\documentclass[preprint,12pt]{elsarticle}
\usepackage{amssymb}
\usepackage{amsmath}

\usepackage[table,xcdraw]{xcolor}
\usepackage[noend]{algpseudocode}
\usepackage{algorithm}
\usepackage{booktabs} 

\usepackage{lmodern}
\usepackage{url}

\journal{Optical Switching and Networking}

\begin{document}

\begin{frontmatter}



\title{Cost and Power-Consumption Analysis for Power Profile Monitoring with Multiple Monitors per Link in Optical Networks}
\tnotetext[label1]{This journal paper is an extension of the conference version, which has been accepted by ECOC 2023.}

% \author[label1]{Qiaolun~Zhang$^{a,b}$, Patricia Layec$^{b}$, Alix~May$^{b}$, %~\IEEEmembership{Senior Member,~IEEE},
% Annalisa~Morea$^{c}$,
% Aryanaz~Attarpour$^{a}$, and~Massimo~Tornatore$^{a}$} %% Author name

% \author[label1,label2]{Qiaolun Zhang}

%% Defining the corresponding author
\author[label1,label2]{Qiaolun Zhang\corref{cor1}}
\cortext[cor1]{Corresponding author}
\ead{qiaolun.zhang@mail.polimi.it}
\author[label2]{Patricia Layec}
\author[label2]{Alix May}
\author[label3]{Annalisa Morea}
\author[label1]{Aryanaz~Attarpour}
\author[label1]{Massimo~Tornatore}

% \affiliation[label1]{organization={},

% \affiliation[label2]{organization={},

\affiliation[label1]{organization={Politecnico di Milano,Department of Electronics, Information and Bioengineering},%Department and Organization
            addressline={Via Ponzio 34}, 
            city={Milan},
            postcode={20133}, 
            country={Italy}}
\affiliation[label2]{organization={Nokia Bell Labs France},%Department and Organization
            addressline={12 Rue Jean Bart}, 
            city={Massy},
            postcode={91300}, 
            country={France}}
% Via Energy Park, 1420871 Vimercate MB, Italy
\affiliation[label3]{organization={Nokia},%Department and Organization
            addressline={Via Energy Park 14}, 
            city={Vimercate},
            postcode={20871}, 
            country={Italy}}


%% Abstract
\begin{abstract}
% Network monitoring is essential to collect comprehensive data on signal quality in optical networks. As deploying large amounts of monitoring equipment results in elevated cost and power consumption, novel low-cost monitoring methods are continuously being investigated. 
As deploying large amounts of monitoring equipment results in elevated cost and power consumption, novel low-cost monitoring methods are being continuously investigated. 
A new technique called \textit{Power Profile Monitoring} (PPM) has recently gained traction thanks to its ability to monitor an entire lightpath using a single post-processing unit at the lightpath receiver. PPM does not require to deploy an individual monitor for each span, as in the traditional monitoring technique using \textit{Optical Time-Domain Reflectometer} (OTDR). 
% PPM and OTDR have different monitoring applications, which will be elaborated in our discussion, hence they can be considered either alternative or complementary techniques according to the targeted monitoring capabilities to be implemented in the network. 
In this work, we aim to quantify the cost and power consumption of PPM (using OTDR as a baseline reference), as this analysis can provide guidelines for the implementation and deployment of PPM. % as either an alternative or a complementary solution of OTDR.} 
First, we discuss how PPM and OTDR monitors are deployed, and we formally state a new Optimized Monitoring Placement (OMP) problem for PPM. Solving the OMP problem allows to identify the minimum number of PPM monitors that guarantees that all links in the networks are monitored by at least $n$ PPM monitors (note that using $n>1$ allows for increased monitoring accuracy). We prove the NP-hardness of the OMP problem and formulate it using an Integer Linear Programming (ILP) model. Finally, we also devise a heuristic algorithm for the OMP problem to scale to larger topologies. 
Our numerical results, obtained on realistic topologies, 
suggest that the cost (and power) of one PPM module should be lower than 2.6 times that of one OTDR for nation-wide and 10.2 times for continental-wide topology. 
% suggest that the cost (power) of one PPM module should be lower than 2.6 times and 10.2 times that of one OTDR for nation-wide and continental-wide topology, respectively. %, which corresponds to 53\% (33\%) of additional cost (power) added to one transponder when PPM is implemented as a submodule of a transponder. 
% 53\% (33\%) that of one transponder. 
\end{abstract}


\end{frontmatter}



\sloppy

\section{Introduction}
\label{sec1}
%% Labels are used to cross-reference an item using \ref command.

For several critical management tasks in optical networks, monitoring equipment is essential to collect comprehensive data on impairments affecting optical transmissions and components. Leveraging data collected by monitors, we can reduce optical margins, detect anomalies for troubleshooting, and ultimately optimize the system performance~\cite{he2023improved,wang2024digital,vejdannik2023leveraging,pang2024large}. Among all the monitored parameters, power has the most significant impact, as it can be used for identifying and localizing power losses, for tracking changes/splices, for fiber characterization, etc. Traditional monitoring techniques mainly retrieve the power using Optical Channel Monitors (OCMs)~\cite{he2023improved,wang2025multi} and Optical Time-Domain Reflectometers (OTDRs)~\cite{zhou2016field}. OCM is typically used to measure the power after amplification in amplifiers, but it cannot be used to obtain the power at each point of the optical line~\cite{he2023improved}. Instead, OTDR injects an \textit{out-of-band signal} to the optical line and is able to estimate the power of the injected signal at each point of the optical line to detect the anomalies in the optical line~\cite{anderson2004troubleshooting}. OTDR's fine power monitoring capability is achieved at the expense of a considerable cost and power consumption, as one OTDR per amplifier is required to monitor the optical line unless additional optical components (such as Loop-Back Line Monitoring System)~\cite{BrianSuboptic} are deployed. Hence, research is ongoing to devise novel low-cost monitoring methods~\cite{Hahn:22,Sasai:23, maySubmine}. 
One of these novel and potentially low-cost methods is the {\em Power Profile Monitoring (PPM)}, which is currently attracting increasing attention~\cite{tanimura2020fiber, may2021receiver, maySubmine}. 
% A new technique, called {\em Power Profile Monitoring (PPM)} is gaining increasing attention~\cite{tanimura2020fiber, may2021receiver, maySubminePower}.
% due to its reduced cost and power consumption.
All PPM solutions estimate the longitudinal power profile through the estimation of the non-linear phase rotation (NLPR) at each point in the link~\cite{sasaiLinear}. Specifically, PPM can monitor an entire LP going through a multi-span optical link by performing a specific digital signal processing (DSP) at the receiver side.  
% The commercial solution to accurately measure the longitudinal \lq\lq loss profile" of a fiber link is the Optical Time Domain Reflectometer (OTDR)~\cite{zhou2016field}. Compared to PPM solutions, unless specifically designed~\cite{BrianSuboptic}, the OTDR is an additional hardware that must be placed at each amplifier, as the OTDR signals do not pass through them. 
% PPM monitors the entire lightpath (LP) by using only an additional post-processing monitoring component at the receiver and, hence, it does not require monitoring each amplifier along the optical line as in traditional monitoring with OTDR~\cite{zhou2016field}. 

The performance of PPM has been significantly improved in the last few years. First, the maximum error on loss has been reduced to 0.25 dB~\cite{KimPPM}, which is comparable to that of OTDR (less than 0.5 dB)~\cite{OTDRref}. 
Second, the spatial resolution (i.e., minimum distance over which two consecutive power events can be distinguished) of PPM has been reduced to 500 meters with measurement distance up to 1200 km~\cite{sasaiLinear}, while the spatial resolution of OTDR is hundreds of meters with a measurement distance of only 80 km~\cite{OTDRref}. The spatial resolution of OTDR can be reduced with a smaller pulse width, but this may result in decreased measurement distance~\cite{OTDRRefRange}. In addition, the measurement distance of PPM has been extended to 10,000 kilometers~\cite{maySubmine}, while the state-of-the-art record of measurement distance of OTDR is up to 28,000 kilometers~\cite{BrianSuboptic}. 
Moreover, to address performance concerns of PPM related to insufficient performance of an LP (due to accumulated noise, such as amplified spontaneous emission or self-phase modulation noises) and an inaccurate transmission distance, Ref.~\cite{mayDemonstration} experimentally demonstrated improving monitoring accuracy by combining estimation from several LPs.

Following recent PPM experimental trials~\cite{maySubmine,KimPPM,sasaiLinear}, 
PPM has confirmed its potential for several monitoring applications, especially considering the limited deployments of OTDR that are often installed only on selected links (especially in backbone networks). % due to its high costs, as each span requires OTDR deployment. 
% especially considering the limited deployments of OTDR in backbone networks. 
Hence, PPM is emerging as a cost-effective and power-efficient monitoring solution supporting various monitoring applications. %such as detecting changes in optical loss, fiber aging, and connector loss, which traditionally depend on OTDR.
Since PPM and OTDR have both \textit{shared and distinct} monitoring applications, they can be considered either \textit{alternative or complementary} techniques according to the targeted monitoring capabilities and budgetary constraints. 
Given this dual role of PPM, it becomes important to quantify the cost and power consumption of PPM on the network scale, which can be compared to that of OTDR as a baseline reference.

\textcolor{black}{PPM technology is still evolving, and its implementation (and hence its cost and power consumption) remains undefined. 
Thus, in this study, we provide a sensitivity analysis across possible cost and power consumption ranges to be used as a guideline for incoming implementations.} 
% %, as a comparison to those of OTDR, 
Considering the ongoing debate on whether PPM should be integrated as a separate module or as a submodule within the receiver, we further evaluate the additional costs and power consumption added to the receiver when integrating PPM into the receiver. 
Our analysis compares the impact of different network architectures, namely, \textit{opaque} vs. \textit{transparent} \lq \lq IP over Wavelength Division Multiplexing'' (IPoWDM) network architectures~\cite{zhang2024power}. Specifically, in transparent architecture, one LP may traverse several links instead of only one link, as in opaque architecture, leading to less required PPMs. Moreover, we evaluate the cost and power consumption required to improve monitoring accuracy using PPM on several LPs.

This is an extended version of our previous work in~\cite{Qiaolun:23}. 
The main contributions and the difference of this work compared to~\cite{Qiaolun:23}  are as follows.
\begin{itemize}
    \item We are the first, to the best of our knowledge, to investigate the problem of optimized monitoring placement (OMP) and quantitatively compare the cost and power consumption of PPM and OTDR. Compared to~\cite{Qiaolun:23}, we extend the OMP problem from only one monitor per link to multiple monitors per link. % (i.e., NPL)
    \item We prove the NP-hardness of the OMP problem and formulated an Integer Linear Programming (ILP) model for it. Moreover, to address the scalability issue of ILP, we propose an efficient heuristic algorithm that can optimize the placement of monitoring components. % under different NPL for opaque and transparent architectures.  
    \item We provide extensive illustrative numerical evaluations to compare the cost and power consumption of PPM to OTDR. This evaluation provides guidelines for a network-wide deployment of PPM that is more cost-effective and energy-efficient than OTDR. These results of PPM are compared to a more resource-efficient deployment approach for OTDR compared to~\cite{Qiaolun:23}, to provide a more compelling baseline.
\end{itemize}

The remainder of the paper is organized as follows: 
Sec.~\ref{sec:system-model} first illustrates the deployment of PPM and OTDR for network monitoring and then provides a formal problem statement for optimized monitoring placement. 
Sec.~\ref{sec:omp-algorithm} describes our proposed optimized monitoring placement algorithm. 
Sec.~\ref{sec:numerical-results} presents and discusses the numerical results, comparing PPM to OTDR. Finally, Sec.~\ref{sec:conclusion} concludes the paper and indicates future work.

\section{System Model for Network Monitoring using PPM and OTDR}
\label{sec:system-model}

\subsection{Application Scenarios of PPM and OTDR}

A comparison of the application scenarios for OTDR and PPM can be found in Table~\ref{tab:applicaition-comparison}. 
Both OTDR and PPM share some critical applications, such as optical loss detection and fiber characterization, owing to some foundational similarities in their working principles. 
Specifically, the applications of OTDR and PPM are both based on analyzing the power profile of monitoring signals (i.e., the injected signal and existing signal in LPs for OTDR and PPM, respectively).

\textcolor{black}{Despite these similarities, PPM and OTDR cater to distinct application scenarios due to differences in their operating principles and assumptions about network equipment and existing LP characteristics. Specifically, PPM assumes that dedicated DSP is performed at the receiver side on the signals of existing LPs and needs as inputs the received signal, its pulse shape, symbol rate and frequency as well as the accumulated chromatic dispersion (CD) to compute the longitudinal power profile along the optical path~\cite{sasaiPerformance,yang2024integrating}. Besides, it is worth noticing that some PPM methods (e.g., correlation-based methods) are sensitive to modulation formats~\cite{sasaiPerformance,may2025accuracy}, and knowing the modulation formats can benefit the selection of proper PPM methods. In contrast, OTDR operates independently of existing LPs by injecting test pulses and analyzing the backscattered signals; thus, it does not require the signal or network-specific information assumed by PPM.} 
% \textcolor{black}{The abovementioned differences in applications between OTDR and PPM are enabled also with different assumptions on capability of network operators. }
\textcolor{black}{
As PPM can estimate the longitudinal power of the LP going through a multi-span optical line, PPM can be used to monitor the parameters of network devices, such as optical filter detuning~\cite{sasaiDigital}. 
By applying PPM to different LP with different wavelength, the gain spectrum of amplifiers can also be estimated~\cite{sena2022advanced}. 
By computing a power profile per polarization, the polarization dependent loss (PDL) can be located and estimated~\cite{may2022receiver,andrenacciPdl}. Other works proposed to use PPM for the monitoring of the physical parameters of the fiber itself. As the x-axis of the estimated power profiles is linked to the accumulated CD, span-wise CD map can be estimated~\cite{sasaiDigital}. Drawing on this, fiber type identification was also demonstrated using PPM~\cite{10484838}. Other physical parameters such as multi-path interference~\cite{hahnLocalization} or longitudinal differential group delay (DGD)~\cite{10527091} can also be estimated. Finally, some works proposed the monitoring of “telecom” parameters, such as the GSNR~\cite{10810103} and the non-linear SNR~\cite{10526939}. 
}
% \textcolor{black}{
% Moreover, since PPM can estimate the power of the LP rather than the injected out-of-band signal as in OTDR, PPM can monitor LP such as span-wise chromatic dispersion (CD) map estimation~\cite{sasaiDigital}, distributed optical power monitoring per wavelength~\cite{sasaiPerformance}, monitoring of polarization-dependent loss~\cite{andrenacciPdl}, multi-path interference~\cite{hahnLocalization}, etc. Specifically, starting from the early works of PPM, it is proposed as approach that can not only obtain power profile, but also extract multiple physical characteristics such as longitudinal fiber losses, CD map, multiple amplifiers’
% gain spectra~\cite{sasaiDigital}. PPM methods can also estimate the power profile at different polarizations, which can be used for localization and estimation of PDL~\cite{may2022receiver,andrenacciPdl}. 
% }

\begin{table}[h]
\small
\centering
\setlength\tabcolsep{3.6pt} 
\caption{Application Comparison between OTDR and PPM.}
\label{tab:applicaition-comparison}
\begin{tabular}{|c|c|c|}
\hline
\rowcolor[HTML]{FFFFFF} 
Application scenarios              & OTDR          & PPM           \\ \hline
\rowcolor[HTML]{FFFFFF} 
Monitoring of change in optical loss             & Available     & Available     \\ \hline
\rowcolor[HTML]{FFFFFF} 
Monitoring of fiber aging/splices                & Available     & Available     \\ \hline
\rowcolor[HTML]{FFFFFF} 
Monitoring of connector loss                     & Available     & Available     \\ \hline
Fiber characterization             & Available     & Available     \\ \hline
\begin{tabular}[c]{@{}c@{}} Monitoring of amplifiers (gain \\ and possible anomalies) \end{tabular}     & Not available & Available \\ \hline
% Monitoring of amplifiers (gain and possible) & Not available & Available     \\ \hline
\begin{tabular}[c]{@{}c@{}} Monitoring of anomaly filters with a \\ detuned center frequency\end{tabular}    & Not available & Available \\ \hline
% \begin{tabular}[c]{@{}c@{}}Spatial and spectral power profile \\ estimation\end{tabular}     & Not available & Available \\ \hline
% \begin{tabular}[c]{@{}c@{}}Spatial and spectral power profile \\ estimation\end{tabular}     & Not available & Available \\ \hline
\begin{tabular}[c]{@{}c@{}} Distributed optical power monitoring \\ per wavelength\end{tabular}     & Not available & Available \\ \hline
% longitudinal differential group delay (DGD)
Monitoring of polarization-dependent loss        & Not available & Available     \\ \hline
\begin{tabular}[c]{@{}c@{}}Span-wise chromatic dispersion (CD)\\ map estimation\end{tabular} & Not available & Available \\ \hline
\begin{tabular}[c]{@{}c@{}}Longitudinal differential group delay (DGD) \\ estimation\end{tabular} & Not available & Available \\ \hline
\begin{tabular}[c]{@{}c@{}}GSNR and non-linear SNR \\ estimation\end{tabular} & Not available & Available \\ \hline
Multi-path interference            & Not available & Available     \\ \hline
Localization of fiber cut          & Available     & Not available \\ \hline
\begin{tabular}[c]{@{}c@{}} Fiber quality assessment before the \\ system is up \end{tabular} & Available & Not available \\ \hline
\end{tabular}
\end{table}

\textcolor{black}{
Conversely, OTDR supports applications not supported by PPM, as it does not require working LPs. Specifically, OTDR can localize fiber cut, while PPM cannot, since no signal is received in the receiver in case of fiber cut. Similarly, OTDR can diagnose fiber quality before the system is up and running by injecting signal, while PPM cannot unless generating test traffic through transponders.}

% This work does not aim to replace OTDR with PPM in those application scenarios that can not be covered by PPM (e.g., localization of fiber cut). 
With this work, it is not suggested to replace OTDR with PPM as some scenarios cannot be covered by PPM (e.g., localization of fiber cut). 
% or some network information (e.g., symbol rate and frequency of the received signal) is not available. 
Instead, if the operators want other monitoring capabilities not covered by OTDR using PPM, our work provides a baseline for evaluating the cost and power consumption of deploying PPM as a complementary monitoring technique. In addition, as a secondary scenario, for networks without OTDR (less likely scenario), when real-time fiber cut localization is not mandatory (note that the localization of links with fiber cut can be determined without deploying OTDR as in Ref.~\cite{delezoideField}), PPM can still offer an alternative to OTDR based on the applications (such as detecting change in optical loss, fiber aging, connector loss, etc.) that are most in the interest of the operator, and that can be provided by PPM.

\subsection{Implementation Options of PPM}

% The three main options for PPM implementation are \textit{(i)} as a separate module outside the receiver, \textit{(ii)} as a submodule inside the receiver, or \textit{(iii)} through offline processing in servers. 
Our study can be applied to three implementation options of PPM, namely, \textit{(i)} as a separate module outside the receiver, \textit{(ii)} as a submodule inside the receiver, or \textit{(iii)} through offline processing in servers. 
If we implement the PPM with the first two options, we need to deploy dedicated network elements, while the third option may use existing servers to do the post-processing. Hence, the cost and power consumption of the first two options are more meaningful to operators. 
Note that while offline server processing eliminates the need for additional hardware and hence has the lowest management cost, it does not support real-time processing and is typically implemented by sending data samples in batches over specified time intervals. Moreover, it incurs high telemetry costs for continuously transmitting data to the server. 
% Instead, this method is typically implemented by sending data samples in batches over specified time intervals. 
% This study focuses on analyzing the extra cost and power consumption associated with enabling \textit{real-time monitoring with PPM}, which is relevant to the first two implementation options. 
For the case of a separate module outside the receiver, the key advantage is its flexibility, as this approach does not impact the transponder's design and allows for easier updates or replacements. In contrast, for the case of implementing PPM as a submodule, the key advantages are related to saving space and weight by integrating the module with the transponder. Besides, this approach could be more power-efficient as it utilizes the transponder's existing resources. 
% However, implementing PPM as a submodule inside the receiver has a higher cost than implementing PPM as a separate module outside the receiver.

Operators can choose the most suitable implementation option by weighing these advantages and disadvantages. However, not all the receivers need to be equipped with PPM because we only need to monitor some LPs among all the LPs traversing a link. Hence, this work aims to optimize the deployment of the PPM modules (either implemented as a separate module or as a submodule of a receiver). 
% This analysis, therefore, is particularly applicable to the options of implementing PPM as either a separate module outside the transponder or as a submodule inside the transponder.

 % \vspace{-1mm}
\begin{figure}[htpb]
   \centering
    \includegraphics[width=0.85\linewidth]{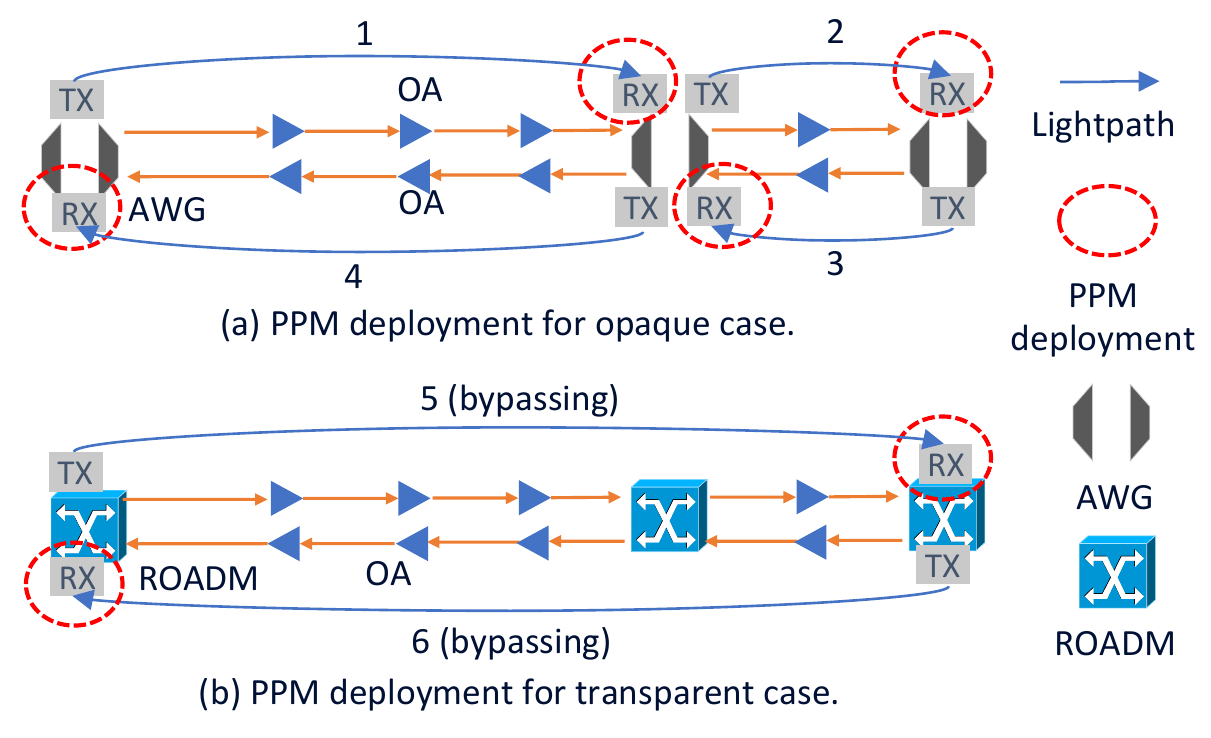}
    \caption{Network-wide PPM deployment.}
    \label{fig:network-architecture}
\end{figure}

\subsection{Deployment of PPM for Network Monitoring}
Fig.~\ref{fig:network-architecture} shows how PPM modules can be deployed for both opaque and transparent IPoWDM network architectures~\cite{zhang2024power}, where we consider a bidirectional network with two fibers per degree. 
Fig.~\ref{fig:network-architecture}(a) depicts the \textbf{opaque} case, where one monitoring module can be placed at each traversed node (except the source node) as opto-electronic conversion is performed at each intermediate node.
% traffic grooming and regeneration
The optical layer uses Arrayed Waveguide Grating (AWG) for muxing and demuxing. In this case, four PPM modules are placed at the receiver of four LPs (i.e., LPs 1-4) to monitor the network. 
Fig.~\ref{fig:network-architecture}(b) depicts the \textbf {transparent} case, where nodes are equipped with Reconfigurable Optical Add/Drop Multiplexer (ROADMs) to enable transparent optical bypass. This architecture allows electrical grooming and traffic regeneration at intermediate nodes with IP routers~\cite{zhang2024power}. In this case, PPM modules can be deployed in each intermediate node with electrical regeneration or at the destination node. As shown in Fig.~\ref{fig:network-architecture}(b), since all the two LPs (LPs 5 and 6) bypass the intermediate nodes, only two PPM modules instead of four PPM modules (opaque case) are required. 
\begin{figure}[htpb]
   \centering
    \includegraphics[width=0.85\linewidth]{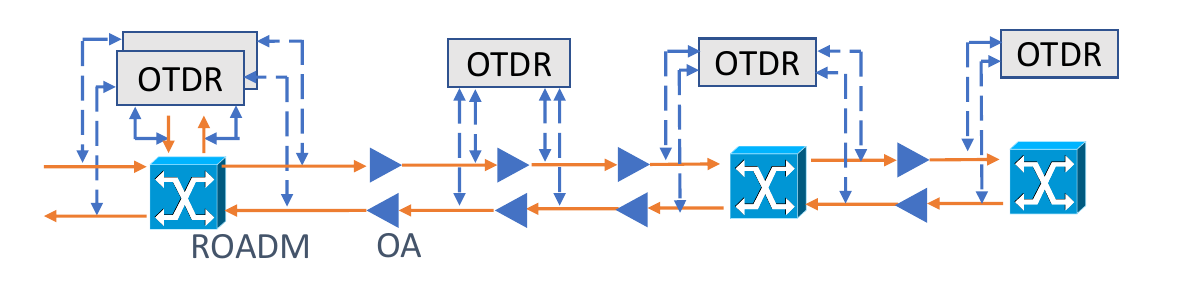}
    \caption{Network-wide OTDR deployment.}
    \label{fig:network-view-OTDR}
\end{figure}
\subsection{Deployment of OTDR for Network Monitoring}

A possible network-wide OTDR deployment for transparent network architecture is shown in Fig.~\ref{fig:network-view-OTDR}, which can also be applied to opaque architecture. 
% In each node, optical channels are switched toward one of the node's output directions or toward transponders (TP) through add/drop block (ADB).
Let us assume the network is denoted by $G=(N,E)$, where $N$ is the set of physical nodes and $E$ is the set of physical links. The number of spans in link $(i,j)$ and the degrees of node $i$ are denoted with $s_{(i,j)}$ and $d_i$, respectively. 
One OTDR card serves 4 fiber links.
% (4 transmitter and 4 receiver ports). 
We assume that no additional optical components except OTDR are used for monitoring, hence OTDRs are needed in every degree of nodes and every inline amplifier (one OTDR can cover both directions as in Fig.~\ref{fig:network-view-OTDR}). Since we assume a bidirectional network with two fibers per degree, the number of OTDRs for node $i$ is $\lceil \frac{2d_{i}}4 \rceil$.
The number of inline amplifiers that need to be monitored in link $(i,j)$ is $max(0, \lceil \frac{s_{(i,j)} - 2}{2} \rceil)$ as the two spans close to both end nodes are monitored with the OTDR for nodes.
Hence, the number of OTDRs required in a network can be calculated using the equation below, which applies to both opaque and transparent architectures:
% \vspace{-1mm}
\begin{equation}
\small
    P = \sum\limits_{(i,j) \in E} max(0, \lceil \frac{s_{(i,j)} - 2}{2} \rceil) + \sum\limits_{i \in N} \lceil \frac{2d_{i}}{4} \rceil
    \label{eq:num-otdr}
\end{equation}
% \vspace{-6mm}

\subsection{\textcolor{black}{Illustrative Example}}
\begin{figure}[htpb]
   \centering
    \includegraphics[width=0.86\linewidth]{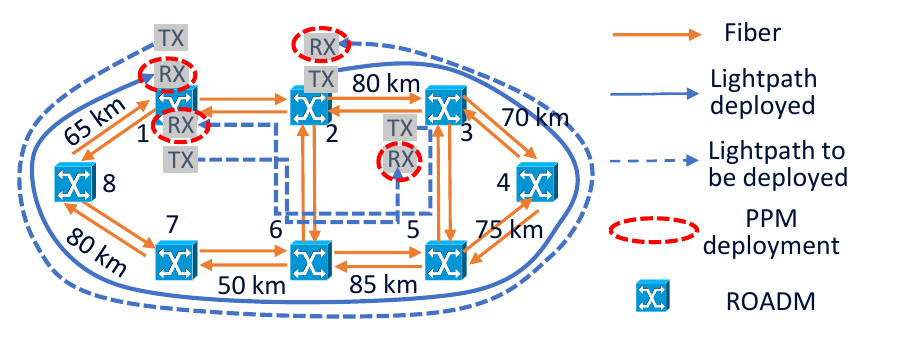}
    \caption{\textcolor{black}{Illustrative Example of Deployment of PPM at Nokia Testbed.}}
    \label{fig:nokia-network-example}
\end{figure}

\textcolor{black}{
We now illustrate network monitoring with OTDR and PPM using an example from our 8-node testbed as in Fig.~\ref{fig:nokia-network-example}. Specifically, we experimentally monitored seven links along the LP marked with solid blue line, from node 2 to node 1, using a single PPM module deployed at node 1 in Ref.~\cite{may2025accuracy}. To monitor the other links, we need to deploy additional PPM modules on other LPs. Assume that we have three additional LPs marked with dashed blue lines as in Fig.~\ref{fig:nokia-network-example}. These LPs can be monitored using three additional PPMs, enabling monitoring of all fibers in the testbed. In contrast, monitoring all fibers with OTDR requires 12 OTDR modules, as calculated by Eqn.~(1). Specifically, in this testbed, each link consists of a single span, and there are no inline amplifiers to monitor. Regarding monitoring of nodes, since one OTDR card can serve 4 fiber links, nodes 1, 4, 7, and 8 require one OTDR module each, while nodes 2, 3, 5, and 6 require two OTDR modules each. In summary, this example demonstrates that to monitor the optical networks, the number of PPMs required is significantly lower than the number of OTDRs.}

 % \vspace{-1mm}

\subsection{Related Work on Monitoring Placement}
Optical performance monitoring has attracted a significant amount of attention~\cite{angelou2012optimized,he2023improved}, as it provides monitoring data with optical monitors for failure localization~\cite{delezoideField,christodoulopoulos2016exploiting,wang2025multi}, swift network recovery~\cite{ZhangProgressive22,arrigoni2023tomography}, optimization of system performance~\cite{he2023improved}, etc. 
% Since ubiquitous monitoring leads to excessive cost, many studies~\cite{machuca2007optimal,angelou2012optimized,monoyios2018attack,he2023improved} have investigated monitoring placement problems to reduce the cost of monitors for mainly two types of applications as follows. 
The study of monitoring placement problems in optical networks dates back to the early 2000's, but it has been revamped more recently following the growing importance of monitoring~\cite{machuca2004optimal,angelou2012optimized,thiruvasagam2021reliable,wang2025multi} for network automation to performance of the network~\cite{pang2024large,ZhangProgressive22,li2024drl,wang2024alarmgpt}. 
% The monitoring performance 
Existing studies focus primarily on reducing the cost of monitors in two scenarios. 
In the first scenario, the objective is to minimize the number of monitors required to determine the faulty (or suspected faulty) network elements~\cite{machuca2004optimal,wang2025multi,machuca2007optimal}. These works are based on machine learning algorithms~\cite{wang2025multi} or heuristic algorithms~\cite{machuca2004optimal,machuca2007optimal}, and work under the assumption that, when an abnormal signal is detected on a monitor, at least one network element before the monitor in the corresponding LP is a faulty element. 
In the second scenario, the objective is again to minimize the number of placed monitors, but in this case, the constraint is to guarantee accurate assessment of some physical metrics (e.g., OSNR, power, etc.) for continuous network optimization (e.g., reduce the network margin)~\cite{angelou2012optimized,thiruvasagam2021reliable}. 
% monoyios2018attack
For instance, Ref.~\cite{angelou2012optimized} minimizes the number of monitors in the network to assess the quality of transmission (QoT) of all the established connections. % Ref.~\cite{monoyios2018attack} investigates monitoring placement to monitor the locations more susceptible to harmful signal interference, which can lead to service degradation. %abrams2004set
Unlike previous optical network monitors~\footnote{Note that we consider only PPM optimized placement, as the number of OTDR cannot be optimized, as discussed in Sec.~\ref{sec:system-model}.1.}, which obtain physical metrics at specific locations, PPM monitors power at each point along the optical line, reducing the number of necessary monitors in an optical line to just one. 
Thus, monitoring placement of PPM is different from previous works on optical network monitoring, and has not been investigated before. Although problems similar to PPM monitoring placement have also been solved in other communication networks (e.g., wireless sensor networks)~\cite{abrams2004set} where one monitor monitors multiple devices/locations, our work still has several key distinctions. Specifically, unlike works for other communication networks~\cite{abrams2004set}, where the potential monitoring placement locations are fixed, our study investigated the monitoring placement of PPM when LPs (i.e., potential monitoring placement locations) vary with traffic loads. Furthermore, we do not only optimize the number of monitors as in Ref.~\cite{abrams2004set}, but also perform a sensitivity analysis to compare the cost and power consumption of PPM to that of OTDR, providing guidelines for future implementations. 

\section{Optimized Monitoring Placement}
\label{sec:omp-algorithm}

In this section, we first formally define the OMP problem. To solve the problem, we formulate an ILP model and propose a scalable heuristic algorithm. 

\subsection{Problem Statement}
This study considers placing PPMs on an initial set of LPs obtained from an existing heuristic algorithm in Ref~\cite{zhang2024power} that minimizes the number of transponders used to serve requests. The reach table of the considered transponder is shown in Table~\ref{tab:modulation-format-long-haul}, which is based on Ref.~\cite{zamiReachTable} and subject to slight modification during internal verification of Nokia. 
To evaluate improving the monitoring accuracy of PPM with several LPs, we defined a new metric called \textit{Number of PPMs per Link} (NPL, defined as the number of monitored LPs that traverse the link). 
% Assume that we want to monitor the links with NPL equals to \textit{k} ({\em nb} a link is monitored with NPL equals to $k$ if at least $k$ LPs traversing it are equipped with one PPM module). 
When the network is in low load, some links may not be monitored with the required NPL, as these links are not traversed by enough LPs. 
% Moreover, the complexity of PPM may increase with the transmission distance as in Ref.~\cite{sasaiDigital}, which imposes constraints on the maximum \textit{monitoring distance} (defined as the maximum distance of the LP that can be monitored by PPM) to reduce cost and power consumption of one single PPM. Thus, the LPs that can be placed with PPMs must ensure the transmission distance is within the maximum monitoring distance.  

% We define \textit{violations of NPL} as the sum of the additional number of LPs required to achieve the required \textit{NPL} of each link in the network. 
% Assume that the number of LPs traversing an edge $e$ to be placed with PPMs is $m_e$ and the required. 

\begin{table}[h]
\small
\centering
\setlength\tabcolsep{3.9pt} 
\caption{Reach table for long-haul transponders.} 
\label{tab:modulation-format-long-haul}
\begin{tabular}{|c|c|c|c|}
\hline
Data rate (Gb/s) & \begin{tabular}[c]{@{}c@{}}Modulation \\ format\end{tabular} & Spacing (GHz) & Reach (km) \\ \hline
800 & PCS 64 QAM & 100 & 150  \\ \hline
700 & PCS 64 QAM & 100 & 400  \\ \hline
600 & 16 QAM     & 100 & 700  \\ \hline
500 & PCS 16 QAM & 100 & 1300 \\ \hline
400 & PCS 16 QAM & 100 & 2500 \\ \hline
300 & PCS 16 QAM & 100 & 4700 \\ \hline
200 & PCS 16 QAM & 100 & 5700 \\ \hline
\end{tabular}
\end{table}

The problem of optimized monitoring placement (OMP) of PPM can be stated as follows: \textbf{Given} a network topology and a set of LPs, \textbf{Decide} the placement of PPM modules, \textbf{Constrained by} placing PPMs over the existing LPs, 
% monitoring of all the links with NPL equals to $k$({\em nb} a link is monitored with NPL equals to $k$ if at least $k$ LPs traversing it are equipped with one PPM module), 
with the \textbf{Objective} of minimizing the \textit{unsatisfied NPL} (defined as the additional number of monitors required to achieve the required NPL) in all links as the first objective and minimizing the number of PPM modules as the second objective. 
% of minimizing the \textit{violations of NPL} as first objective and minimizing the number of PPM modules as second objective. 

\subsection{NP-hardness of the OMP Problem}

% In this section, we 
We prove that the OMP problem is NP-hard when all the links can achieve the required NPL with Theorem 1 as follows. 
% Theorem 1 indicates the proof of NP-hardness of the OMP problem.
%We now prove that the OMP problem, when all the links can achieve the required NPL, is NP-hard. 
%The proof of the NP-hardness of the OMP problem is shown in Theorem 1 as follows.

\textit{Theorem 1:}
The OMP problem is NP-hard when all the links can achieve the required NPL (i.e., there are enough LPs such that all the links can be monitored with the required number of monitors, denoted with $\gamma$).

\textit{Proof:}
First, we show that a simplified version of OMP, where the number of LPs that exclusively traverse all edges in each physical path $l \in L$ is at most 1, is NP-hard. Since all links can achieve the required NPL, all links are monitored with $\gamma$ PPMs after satisfying the primary objective. Then, the simplified OMP problem (i.e., OMP problem of satisfying the second objective) can be stated as follows: \textbf{Given} a set of links and a set of LPs, \textbf{Decide} the placement of PPM modules, \textbf{Constrained by} satisfying the required NPL of all links, with the \textbf{Objective} of minimizing the number PPM modules. We show as follows that the \textit{set k-cover problem}~\cite{slijepcevic2001power}, a known NP-hard problem, can be reduced to the decision version of the simplified OMP problem.

The set k-cover problem can be stated as follows: Given a set $E$ of $|E|$ elements, a collection $S={S_1, S_2, ..., S_m}$ of $m$ subsets of $E$, parameter $\delta_l^e$ of whether subset $S_l$ contains element $e$, an integer $\gamma$, and an integer $v$, does there exist a set of no more than $v$ subsets (defined as $I = \{S_1, ..., S_v\}	\subseteq S$) such that $\sum\limits_{S_l \in I} \delta_l^e \geq \gamma$ for all $e \in E$. 
% $\sum\limits_{S_l \in I} \sum\limits_{e \in S_l} \delta_l^e \geq \gamma$.  
Given an instance of the set k-cover problem, an instance of the decision version of the simplified version of the OMP problem can be constructed as follows. 
Assume that the set of edges is denoted with $E$, and the set of traversed edges of physical path $l$ is denoted with $S_l$. 
% We use $E$ to define the set of edges, and set $S_l$ as the set of traversed edges of physical path $l$. 
The parameter $\delta_l^e$ corresponds to whether a physical path $l$ traverses edge $e$. The integer $\gamma$ denotes the required NPL. The corresponding question becomes whether a solution to the simplified OMP problem exists, such that the number of selected physical paths to monitor with PPMs is no more than $v$.  The simplified OMP problem reaches the optimum if and only if the set k-cover problem reaches the optimum. In conclusion, the OMP problem is NP-hard when all links can achieve the required NPL. 

% an integer $V$ and a subset $\overline{S}$ of set $S$, such that $\sum\limits_{S_l \in \overline{S}} \sum\limits_{e \in E} \delta_l^e \geq \gamma$ for all $e \in E$, and 
% Given a set of physical paths $L$, a set of links $E$, the cost of placing one PPM over the LP exclusively traversing all edges in physical path $l$ (defined as $\beta$), 

\subsection{Integer Linear Programming Model for OMP Problem}
The sets and parameters for the ILP model are shown in Table~\ref{tab:parameters}, and the variables are shown in Table~\ref{tab:variables}. 

\begin{table}[htbp] 
\small
\centering
\setlength\tabcolsep{5pt} 
	% \captionsetup{font=footnotesize}
	\caption{Sets and parameters for the ILP Model}
	\label{tab:parameters}
	\begin{tabular}{ll} 
		\toprule 
	  Notation & Description\\
		\midrule 
		% $N$ & Set of physical nodes\\
		$E$ & Set of physical links \\
        $L$ & Set of physical paths \\
        $c_l$ & Number of LPs that exclusively traverse all edges \\
        & in physical path $l \in L$\\
        $\delta_l^e$ & Equals to 1 if physical path $l \in L$ traverses $e \in E$ \\
        $\gamma$ & Required NPL \\
        $\alpha$ & Weight of number of unsatisfied of PPMs \\
        % $p_l$ & Integer, number of LPs traversing physical path \\
        % & $l \in L$ selected for placing PPMs\\
        % $x_e$ & Integer, number of PPMs for $e \in E$ \\
        % % $y_$
		\bottomrule 
	\end{tabular} 
\end{table}
% \vspace{-5mm}
\begin{table}[htbp] 
\small
\centering
\setlength\tabcolsep{8.2pt} 
	% \captionsetup{font=footnotesize}
	\caption{Variables for the ILP Model}
	\label{tab:variables}
	\begin{tabular}{ll} 
		\toprule 
	  Notation & Description\\
		\midrule 
        $p_l$ & Integer, number of monitored LPs exclusively  \\
        &  traversing all edges in  physical path  $l \in L$ \\
        $x_e$ & Integer, number of PPMs for $e \in E$ \\
        % $y_$
		\bottomrule 
	\end{tabular} 
\end{table}

The objective function of the ILP model is shown in Eqn.~(\ref{eq:goal}), where $\sum\nolimits_{e \in E} (\gamma - x_e)$ denotes the sum of unsatisfied NPL in all links and $\sum\nolimits_{l \in L} p_l$ denotes the number of PPMs used. 
% where the objective of maximizing the sum of achieved NPL in all links (i.e., $\sum\nolimits_{e \in E} x_e$) is converted to minimizing the sum of additional number monitors required to achieve the required NPL (i.e., $\sum\nolimits_{e \in E} (\gamma - x_e)$). 
The weight $\alpha$ is set to be greater than the maximum number of hops of all physical paths in the network to prioritize minimizing unsatisfied NPL rather than reducing the PPM modules used. 

The constraints of the ILP model are listed in Eqn.~(\ref{eq:maximum_ppm_per_link})-Eqn.~(\ref{eq:number-selected-path}). Eqn.~(\ref{eq:maximum_ppm_per_link}) ensures the upper bound of the achieved NPLs for all links. 
Eqn.~(\ref{eq:num_ppm_for_link}) ensures that the number of PPMs for link $e \in E$ is less or equal to the number of monitored LPs that traverse link $e$. 
Eqn.~(\ref{eq:number-selected-path}) ensures that the number of monitored LPs exclusively traversing all edges in physical path $l \in L$ should be smaller or equal to the number of available LPs. 
% number of LPs that exclusively traverse all edges in the physical path $l \in L$. 
\begin{equation}
	%\begin{split}
	\label{eq:goal}
	\min \hspace{10pt} \alpha \sum\limits_{e \in E} (\gamma-x_e) + \sum\limits_{l \in L} p_l
	%\end{split}
\end{equation}

\begin{equation}
    x_e \leq \gamma \quad \forall e \in E
    \label{eq:maximum_ppm_per_link}
\end{equation}
\begin{equation}
    x_e \leq \sum\limits_{l \in L} \delta_l^e p_l \quad \forall e \in E
    \label{eq:num_ppm_for_link}
\end{equation}
\begin{equation}
    p_l \leq c_l \quad \forall l \in L
    \label{eq:number-selected-path}
\end{equation}

\subsection{Optimized Monitoring Placement Algorithm}

We developed a new \textit{OMP} algorithm to reduce the number of PPMs for monitoring. 
The \textit{OMP} algorithm is designed leveraging existing effective heuristic algorithms for the set covering problem as in Ref.~\cite{vasko2016best}. 
Note that this problem is different from the failure localization problem, as the OMP problem aims to \textit{deploy monitoring modules}, while the failure localization problem~\cite{delezoideField,christodoulopoulos2016exploiting} aims to \textit{localize the soft/hard failure} in the network based on the monitoring data.

% \begin{figure}[htpb]
%    \centering
%     \includegraphics[width=1.0\linewidth]{4-comparison-problem.pdf}
%     \caption{Network-wide OTDR deployment.}
%     \label{fig:network-view-OTDR}
% \end{figure}

% \begin{algorithm}[htb]
\begin{algorithm}[!h]
\small
     \caption{Optimized monitoring placement algorithm.}
    \begin{algorithmic}[1]
% \DontPrintSemicolon
% \State \textbf{Input:} $a, b$ \Comment{The inputs of the algorithm}
\State \textbf{Input:} Network topology $G=(N,E)$, $L$, $c_l$, NPL $\gamma$, $d_{max}$
\State \textbf{Output:} Placement of PPM
     % \State Initialize $L_c \leftarrow \{l \in L|d_l \leq d_{max}\}$, $L_m \leftarrow L_c$, $E_m \leftarrow E$
     \State Initialize $L_m \leftarrow L$, $E_m \leftarrow E$
     \State Initialize covering matrix $M$ with $|E|$ rows and $|L|$ columns
     \For{$e \in E$}
        \State $x_e \leftarrow 0, z_e = \sum\limits_{l \in L} c_l M_{(e,l)}$
     \EndFor
     \If{\textit{architecture is opaque}}
        \For{\textit{each} $l \in L$}
        % \State Select any $l$ from $L$
        \State $x_e \leftarrow min(c_l,\gamma)$%, $L_m \leftarrow L_m \setminus \{ {l} \}$
        % \State Continue
        \EndFor
        \State Return
     \EndIf
     \While{$|E_m| > 0$}
        \If{All elements in $M$ is empty}
            \State Break
        \EndIf
        % \State Find the set of $L_s$ such that $L_s = \{l| l \in L \wedge cost_l^1 \geq \}$
        \State Obtain set $L_s$ such that $v_{l^*} \geq max\{ v_l | l \in L_m \}, \forall l^* \in L_s$
        % \State Find the set of $L_s$ such that $\sum\limits_{e \in E} M_{(e,l^*)}=max\{ \sum\limits_{e \in E} M_{(e,l)} | l \in L_m \}, \forall l^* \in L_s$
        
        \State $l_s \leftarrow -1, cost_{l_s} \leftarrow Inf$
        \For{$l \in L_s$}
            \State Obtain $cost_l$ for LP $l$
            \If{$cost_l < cost_{l_s}$}
                \State $l_s \leftarrow l, cost_{l_s} \leftarrow cost_{l}$
            \EndIf
        \EndFor
        
        \For{$e$ \textit{in traversed edges of LP} $l_s$}
            \State $x_e \leftarrow x_e + 1$, $z_e \leftarrow z_e - 1$
            \If{$x_e \geq \gamma$}
                \State Let $M_{(e,l_s)} \leftarrow 0, \forall l_s \in L_m$
                \State Let $E_m \leftarrow E_m \setminus \{e\}$, $z_e \leftarrow 0$
            \EndIf
        \EndFor
        \If{$p_{l_s} \geq c_{l_s}$}
            \State Let $L_m \leftarrow L_m \setminus \{ {l_s} \}$
            \State Let $M_{(e,l_s)} \leftarrow 0, \forall e \in E$
        \EndIf
      \EndWhile
    \end{algorithmic}
    \label{alg:omp}
\end{algorithm}

In the following, we illustrate the details of the proposed \textit{OMP} algorithm through Algorithm.~\ref{alg:omp}. 
The input $L$ denotes the set of physical paths, and the input $c_l$ denotes the number of LPs that traverse a path $l \in L$. 
The inputs $\gamma$ denote the required \textit{NPL}. 
% The inputs $\gamma$ and $d_{max}$ denote the required \textit{NPL} and the maximum monitoring distance of PPM, respectively. 

The algorithm first initializes the sets and variables. Specifically, 
% the set $L_c$ is initialized with all the physical paths whose transmission distance is within the maximum monitoring distance of PPM (line 3). In addition, 
set $L_m$ denotes the set of physical paths that have unmonitored LPs exclusively traversing it, and set $E_m$ denotes the set of links that do not satisfy the required \textit{NPL}.
Then, the algorithm initializes the covering matrix $M$, where each column corresponds to a physical path and has $|E|$ elements (line 4). 
An element in the matrix $M$ equals one when the corresponding column represents a physical path containing the corresponding edge, and there are unmonitored LPs traversing this physical path. 
% An element in the matrix $M$ equals one when the physical path, corresponding to the column of that element, contains the corresponding edge, and there are unmonitored LPs exclusively traversing the physical path. 
% the LP corresponding to the column of the element contains the corresponding edge. 
% Then the algorithm initializes the achieved \textit{NPL} of each link (denoted with $m_e$) as 1 (line 3-4), and the number of LPs that traverse physical path $l$ and are placed with PPMs (denoted with $p_l$) (line 5-6). 
Assume that $z_e$ denotes the number of unmonitored LPs that traverse $e$ when $e$ does not satisfy the required NPL. Otherwise, $z_e$ equals to 0. 
After initializing $M$, the algorithm sets the initial value of the achieved \textit{NPL} for each link (denoted as $x_e$) and $z_e$ (lines 5-6). 
% The algorithm initializes $z_e$ with a
% Hence, $z_e$ denotes the number of unmonitored LPs that can be used to increase the NPL of links. 

Then, the algorithm decides the PPM placement according to the network architecture (line 7-27). 
If the network architecture is opaque, all the LPs have only one hop, and the placement of PPM on LPs that exclusively traverse a physical path does not affect the placement of PPM on LPs that exclusively traverse \textit{any other physical path}.
Thus, the algorithm can simply loop over all path $l \in L$ and decide the number of PPMs placed on the LPs that exclusively traverse $l$ (line 7-10). Instead, for transparent architecture, since different physical paths may have overlapping links, the algorithm decides the placement of PPM until all the links are monitored as follows (line 11-27). 
If no LP can monitor more edges, the algorithm stops monitoring links (line 12-13). 
Assume that the number of links that can be monitored with LP exclusively traversing physical path $l$ is defined as $v_l$ as in Eqn. (\ref{eq:heuristic-cost-1}). The algorithm first greedily finds the set of paths that can be used to monitor the maximum number of links (line 14).  Specifically, the algorithm first obtains the set $L_s$, where each path $l^* \in L_s$ has the maximum $v_l$ among all the paths that can be monitored. 
Then, the algorithm tries to break the ties of the paths in $L_s$ with the cost of monitoring the LP (15-19). 
%that exclusively traverses the physical path $l$ (line 13-17). 
The cost of monitoring an LP is defined as $cost_l$ and can be calculated with Eqn.~(\ref{eq:heuristic-cost-2-z}) and Eqn.~(\ref{eq:heuristic-cost-2}). 
Specifically, as in Ref.~\cite{vasko2016best}, our algorithm gives priority to monitoring the edges that are covered by less LPs. 
The number of unmonitored LPs that can be used to monitor an edge $e$ is defined as in Eqn.~(\ref{eq:heuristic-cost-2-z}), and the cost of placing a PPM in an LP is defined to be positively correlated to the number of unmonitored LPs that can be used to monitor all edges in LP as in Eqn.~(\ref{eq:heuristic-cost-2}). 
After selecting the LP that traverses path $l_s$ to monitor, the algorithm updates NPL for the link (i.e., $x_e$) and $z_e$ (line 21). 
If the NPL of a link achieves the required NPL, the corresponding row of the link in $M$ is set to 0, indicating that the link does not need more monitors (line 23). In addition, link $e$ is removed from the set of links that need to be monitored, and $z_e$ will be set to 0 (line 24). 
At last, if all the LPs that exclusively traverse path $l_s$ are monitored, path $l_s$ will be removed from the set of paths that can be placed with monitors (line 26), and the corresponding column of the path in $M$ is set to 0 (line 27). 
% Then, the algorithm updates the \textit{NPL} for the links in the selected path $l_s$. If an edge satisfies the required $NPL_s$, the corresponding row in $M$ is set to 0 (line 29), and the set $E_m$ is updated. 
% At last, if the number of LPs traversing path $l$ to be monitored is greater or equal to the available number of LPs traversing path $l$, the column corresponding to $l$ is set to 0 to avoid selecting such a path. 
\begin{equation}
    v_l = \sum\limits_{e \in E_l} M_{(e,l)} \quad \forall l \in L
    \label{eq:heuristic-cost-1}
\end{equation}
\begin{equation}
    z_e = \sum\limits_{l \in L} (c_l-p_l) M_{(e,l)} \quad \forall e \in E
    \label{eq:heuristic-cost-2-z}
\end{equation}
\begin{equation}
    cost_l = \frac{1}{\sum\limits_{e \in E_l} 1/z_e}%\frac{1}{z_e}} \quad \forall l \in L
    \label{eq:heuristic-cost-2}
\end{equation}

\begin{figure}[htpb]
   \centering
    \includegraphics[width=0.75\linewidth]{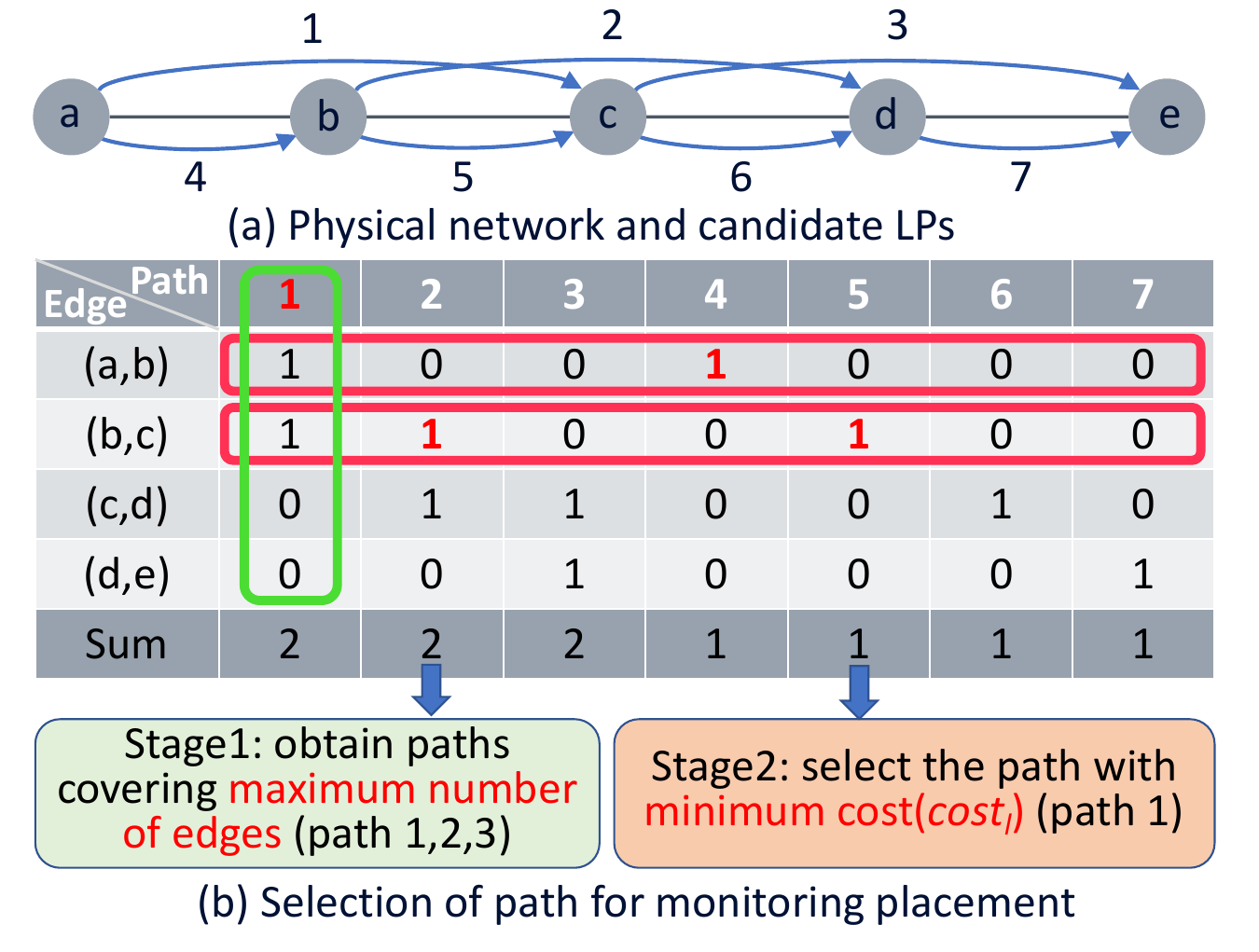}
    \caption{Illustration of the \textit{OMP} algorithm.}
   \label{fig:procedure-algorithm}
\end{figure}

The \textit{OMP} algorithm is illustrated in Fig.~\ref{fig:procedure-algorithm}, where we consider only one direction, for the sake of simplicity. 
Fig.~\ref{fig:procedure-algorithm}(a) shows a simple physical network and seven candidate LPs. No LP is longer than two hops, e.g., due to the limited reach of transponders. Assume that the required NPL is 1. 
% Starting from an empty list of LPs, \textit{OMP} iteratively adds a new LP to the list, by choosing greedily the LP that can monitor more links until all the links are monitored as shown in Fig.~\ref{fig:procedure-algorithm}(b). 
The covering matrix $M$ is shown in Fig.~\ref{fig:procedure-algorithm}(b), representing whether additional edges can be monitored when selecting a path. 
% we can monitor additional edges with a path. 
% , where an element is equal to one when there are unmonitored LPs corresponding to the column of that element, containing the corresponding edge. 
The algorithm iteratively selects new LPs to monitor until all the links are monitored. At each iteration, the algorithm requires two stages. 
% to select the new LP to monitor. 
In stage 1, the algorithm identifies the LPs that allow to cover (monitor) the maximum number of edges, i.e., it selects the columns with the maximum $v_l$ as defined in Eqn.~(\ref{eq:heuristic-cost-1}) (listed in the last row of the table in Fig.~\ref{fig:procedure-algorithm}(b)). In stage 2, the algorithm selects, among the columns identified in stage 1, the column (i.e., the physical path) that has a minimum value of $cost_l$ as in Eqn.~(\ref{eq:heuristic-cost-2}). 
% (i.e., the LP) that affects the minimum number of other LPs (an edge of another LP is affected if both LPs have an element equal to 1 in the same row). 
In case of ties, \textit{OMP} randomly selects a column. 
In Fig.~\ref{fig:procedure-algorithm}(b), after selecting LPs 1, 2, and 3 in phase 1, stage 2 calculates the cost of selecting LPs. Specifically, LP 1 and LP 3 have the smallest cost of 6/5 by calculating the cost with all the traversed edges (marked with red rows for LP 1 in Fig.~\ref{fig:procedure-algorithm})(b) using Eqn.~(\ref{eq:heuristic-cost-2}). 
In the current iteration, the algorithm selects LP1. 
In the next iteration, the algorithm will select LP 3, and all the edges will be monitored. In summary, \textit{OMP} reduces the PPM modules from 7 (one for each LP) to 2.
% Specifically, the cost of selecting LP 1 can be calculated with the $z_e$ of all the traversed edge $e$ of LP 1 with Eqn.~(\ref{eq:heuristic-cost-2}). In this example, the cost of selecting 
% as it will affect the lower number of other LPs. 
% The affected edges of the other paths are marked in red in Fig.~\ref{fig:procedure-algorithm} (b).
% Choosing a particular LP, such as LP 1, affects the number of additional edges that can be monitored when selecting alternative LPs since certain edges may already be monitored in LP 1.
% If we select LP 2, it will affect the number of additional LPs that we can monitor when selecting the other LPs. For instance, when LP 2 is selected, the LPs that have the same links as LP 2 can only monitor fewer edges. 
% The affected edges of the other LPs are marked with red rows and red font in Fig.~\ref{fig:procedure-algorithm}. We identify that LP 1 may affect LPs 2, 4, and 5. As all edges in LPs 4 and 5 are already monitored and therefore will not be selected, only one LP is considered affected. In the current iteration, the algorithm can select LP 1, which only affects one LP. In the next iteration, the algorithm will select LP 3, and all the edges will be monitored. In summary, \textit{OMP} reduces the PPM modules from 9 (one for each LP) to 2.

% \vspace{-3.5mm}
% \vspace{-16pt}

\emph{Complexity of the algorithm:} For opaque architecture, the algorithm iterates all physical paths (at most $|L|$) and decides the number of PPMs with the $min(c_l, \gamma)$ operation, the complexity is $O(|L|)$. Instead, for transparent architectures, the complexity of the algorithm is mainly determined by the process of selecting the LPs to monitor (line 11-27). The number of iterations of selecting the LPs is at most $O(\gamma |E|$). 
Assume that the maximum number of hops of LPs is $H$. The complexity of selecting the set $L_s$ is $O(H|L|)$. The complexity of calculating $cost_l$ is $H$, and hence the complexity of breaking ties of the paths in $L_s$ is $O(H|L_s|)$. The complexity of updating the matrix $M$ after selecting the LP to monitor is $O(H|L|)$. Hence, the overall complexity is $O(\gamma |E| H |L|)$.

\subsection{\textcolor{black}{Extension of the Proposed Solution to Changing Traffic Patterns}}
\textcolor{black}{The current study focuses on static traffic scenarios to evaluate the performance of the proposed heuristic and monitoring strategy. Both our ILP and heuristic algorithm can be extended to dynamic scenarios by periodically re-evaluating the monitoring configuration in response to traffic changes. Specifically, if the PPM modules are implemented through offline processing in servers, changes in traffic patterns can be handled by collecting information on all available LPs and re-computing the network-wide optimized monitor placement. If the PPM modules are implemented as separate devices outside the receiver or as submodules within the receiver, our solution can be extended assuming that  new monitors can only be deployed in addition to the existing ones. 
For the ILP formulation, let $\overline{L}$ denote the set of LPs previously monitored with PPMs. We introduce Eqn.~(\ref{eq:keep-previous-monitors}) to ensure that previously monitored LPs continue to be monitored in future reconfigurations. For the heuristic algorithm, this can be achieved by initializing the algorithm with the previously monitored LPs already marked as covered by PPMs.
}

\begin{equation}
    \textcolor{black}{
    p_l = 1 \quad \forall l \in \overline{L}
    }
    \label{eq:keep-previous-monitors}
\end{equation}

\section{Illustrative Numerical Results}
\label{sec:numerical-results}
\subsection{Simulation Settings}
We implemented both the ILP formulation and the \textit{OMP} algorithm using Python. The ILP formulation was solved with a state-of-the-art non-commercial solver, namely Solving Constraint Integer Programs (SCIP) solver~\cite{bestuzheva2023enabling}. The simulations are performed on a workstation with Intel(R) Core(TM) i5-13400F CPU processor and 32 GB of memory. 

We first compare the ILP and the \textit{OMP} algorithm in terms of the unsatisfied NPL and execution time over Gabriel graph topologies~\cite{gabriel1969new} with nodes ranging from 100 to 300. Gabriel graphs have been shown to accurately reflect characteristics of optical networks~\cite{velinska2017optical}. 
Specifically, the Gabriel Graph models the network topology as a grid-like structure, which closely matches some existing American optical networks, such as AT\&T, Level 3, and Sprint. 
% cetinkaya2014comparative
Then, we perform our numerical evaluations on both a national-scale and a continental-scale real network topology, the 14-node Japan (J14)~\cite{ibrahimi2021machine} and the 14-node NSF (N14), respectively~\cite{bathula2009constraint}. Specifically, we first evaluate the number of monitors under different loads, and then compare the cost and power consumption of PPM and OTDR. 
Note that we evaluate various static scenarios with different loads instead of a dynamic scenario. we generate and test the performance of different placement solutions across these loads to demonstrate that our results can be generalized to real-world scenarios with different traffic loads. 
Since ILP is scalable on both J14 and N14, all the evaluations on J14 and N14 are based on the optimal solutions of ILP. 
In addition, each fiber operates on a 6-THz C-band. 

We consider PPM deployment in two network architectures, i.e., opaque (the cases with and without \textit{OMP} are named as \textit{Op-O} and \textit{Op}, respectively) and transparent (the cases with and without \textit{OMP} are named as \textit{Tr-O}, \textit{Tr}, respectively) architectures, which are compared to the OTDR deployment. 
Note that we only compare PPM to OTDR, as OTDR is the most typical monitors that can monitor the power at each point in optical line as PPM. 
When evaluating the number of monitors under different loads, some links are not crossed by any LP when network traffic is exceptionally low (less than 30 Tb/s). In this case, we do not deploy OTDR on links not traversed by any LPs. 
Since the LPs may vary between opaque and transparent architectures, the links not traversed by any LPs could also differ, potentially leading to a variation in the number of OTDRs deployed in different architectures. 
Note that the difference in the number of OTDRs for opaque and transparent architectures is less than 3\% for all the tested cases only under very low traffic. 
Moreover, when comparing the cost and power consumption under 1\% rejection rate, all the links are traversed by at least one LP due to higher traffic, leading to the same number of OTDRs deployed for both opaque and transparent architectures. 
Thus, we only show the results of OTDR with transparent network architectures for simplicity. 
% the number of OTDRs deployed for opaque and transparent architectures differs solely when network traffic is exceptionally low (less than 30 Tb/s), and there are insufficient LPs to monitor every link. 
% Moreover, the difference in the number of OTDRs for opaque and transparent architectures is less than 3\% for all the tested cases. 
% (note that the OTDR deployment for opaque and transparent architectures is the same when comparing the cost and power consumption of PPM and OTDR, as the evaluation considers monitoring all the links). In addition, 
% Even when the network traffic is lower than 30 Tb/s, the difference in the number of OTDRs for opaque and transparent architectures is less than 3\% for all the tested cases. 
% only under very low network traffic (less than 30 Tb/s and there are not enough LPs to monitor not all the links), 
% and the difference in the number of OTDRs for opaque and transparent architectures is less than 3\% for all the tested cases. 
% When considering a different NPL, 
When considering multiple NPLs, the cases with \textit{OMP} and links with $\gamma$ monitors are named as \textit{Op-O-}$\gamma$ and \textit{Tr-O-}$\gamma$, for opaque and transparent architectures, respectively. Note that, without \textit{OMP}, since we place one PPM in each receiver, the number of placed PPMs is a constant value regardless of the required NPL, and hence we do not need to add suffix \textit{-k} for the cases without \textit{OMP} (i.e., \textit{Op} and \textit{Tr}). 
% Moreover, the cases that enforce the maximum monitoring distance of PPM within \textit{k} kilometers are denoted as \textit{Tr-O(k)} for transparent architectures. To monitor all the links, the enforced maximum monitoring distance should be larger or equal to the maximum link length in the topology, and hence, the enforced maximum monitoring distance does not affect the PPM placement in the opaque architecture. 
For comparison between ILP and the \textit{OMP} algorithm, we assume the transmission distance of long-haul transponders can be up to six hops in the network. Instead, for cost and power consumption analysis with N14 and J14, we use the distance in the real network. 
The requested data rate is randomly generated from 100Gb/s to 400Gb/s among all node pairs. 
% requests are generated between node pairs that have less than six hops and the transmission distance of 
We first increase the number of requests and evaluate the number of monitors under different loads. 
Then, we increase the number of requests until 1\% of requests are blocked in transparent architecture, and we use the corresponding carried traffic to evaluate the cost and power consumption for both opaque and transparent architectures (note that opaque architecture can carry more traffic than transparent architecture). 
% , and the number of requests is set to have 1\% blocking probability for transparent architecture, which can carry less traffic than opaque architecture. 
The normalized cost (energy) of a transponder~\cite{zamiReachTable} and an OTDR module is 4 (8) and 0.2 (0.25), respectively, which is obtained and verified internally in Nokia. 
\textcolor{black}{Current PPM methods are based on solving the inverse problem of the nonlinear Schrödinger equation (NLSE), and their computational complexity has been verified by our Nokia engineers internally to be significantly lower than that of a transponder. In our simulations, the sensitivity analysis of PPM is conducted across a range of up to 60\% of the cost and power of transponders, which serves as an upper bound in our analysis.} 
% The normalized cost (energy) of a transponder and an OTDR module is 4 (8) and 0.2 (0.25), respectively. 
% We vary the cost and power consumption of a PPM module compared to that of a transponder.
% and compare the cost of deploying PPM and OTDR in the network.
All the reported results are averaged over 10 instances. 

\begin{figure}[htpb]
   \centering
        \includegraphics[width=0.7\linewidth]{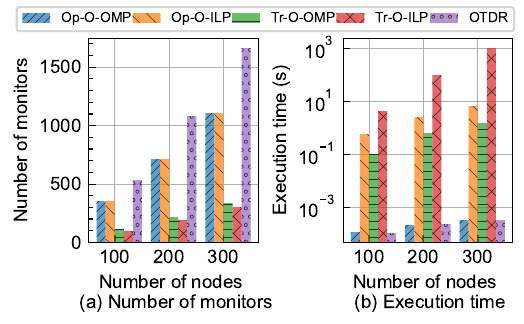}
    \caption{Comparison of the \textit{OMP} algorithm and ILP.}
    \label{fig:comparison-ilp-heuristic}
\end{figure}

\subsection{Comparison of ILP and Heuristic Algorithm}
In this section, we first compare the number of monitors and the execution time using ILP and the heuristic algorithm. The suffix \textit{OMP} and \textit{ILP} denote the method to solve the scenario. Fig.~\ref{fig:comparison-ilp-heuristic}(a) shows the number of monitors under different numbers of network nodes. 
Note that we present results only for topologies with 100 to 300 nodes, as the ILP cannot scale to topologies with 400 nodes. 
First, employing PPM significantly reduces the number of monitors required. In particular, in transparent architectures, the integration of PPM allows for a reduction of up to approximately 83\% in the number of monitors using OTDR. 
% all the scenarios with PPM reduce the number of monitors significantly. Notably, the number of monitors with OTDR can be reduced up to around 83\% with PPM in transparent architectures (i.e., Tr-O-ILP). 
Second
% First
, in the transparent architecture, the number of monitors is reduced up to 74\% with respect to the opaque architecture, since LPs in transparent architecture have a larger number of hops due to bypassing. Regarding the gap between the \textit{OMP} algorithm and the ILP, for opaque architecture, the optimality gap is 0\% for all the tested cases. Instead, for transparent architecture, the optimality gap can be up to around 14.9\%, because different LPs may partially overlap, and it is much more difficult to place PPMs for transparent architecture compared to that for opaque architecture. The performance of the OMP algorithm is achieved with a significant reduction of execution time, e.g., the execution time of transparent architecture reduces from about 4345 seconds to 1 second, as shown in Fig.~\ref{fig:comparison-ilp-heuristic}(b).

\subsection{Evaluation of Number of Monitors under Different Loads}
In the following, we first evaluate the average unsatisfied NPL when increasing the traffic, and then we compare the number of monitors used.

\begin{figure}[htpb]
   \centering
        \includegraphics[width=0.75\linewidth]{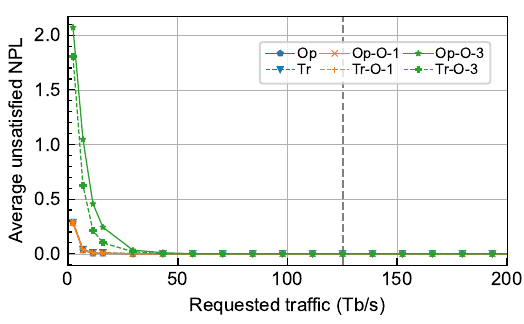}
    \caption{Average unsatisfied NPL vs. load in N14.}
    \label{fig:unsatisfied-NPL-nsf}
\end{figure}

% Let us first evaluate the average unsatisfied NPL in links as in Fig.~\ref{fig:unsatisfied-NPL-nsf}. 
The unsatisfied NPL is reported in Fig.~\ref{fig:unsatisfied-NPL-nsf} where the solid dotted line represents the requested traffic under 1\% rejection rate. 
% in Fig.~\ref{fig:monitors-vs-load-nsf} and Fig.~\ref{fig:monitors-vs-load-japan} represents the requested traffic under 1\% rejection rate. 
When the requested traffic is less than 43Tb/s, the average unsatisfied NPL reduces when the requested traffic increases. This is because, with higher traffic, more LPs can be established, which can be placed with PPMs for monitoring. 
Note that for \textit{Op} and \textit{Tr}, since we do not optimize the achieved NPL, we consider that a link satisfies the required NPL if it is monitored by one PPM. Specifically, for \textit{Op} and \textit{Tr}, the unsatisfied NPL of a link equals 0 if the link has at least one monitor. Otherwise, the unsatisfied NPL of a link equals 1. 
Thus, the unsatisfied NPL of \textit{Op} is the same as \textit{Op-O-1}, and the unsatisfied NPL of \textit{Tr} is the same as \textit{Tr-O-1}. In addition, the lines representing \textit{Op-O-1} and \textit{Tr-O-1} almost overlap because the maximum difference between the unsatisfied NPL of \textit{Op-O-1} and \textit{Tr-O-1} is less than 0.005, indicating the number of links that are monitored with at least one PPM for opaque and transparent architectures are almost the same. 
Instead, with a low load, \textit{Tr-O-3} reduces up to around 100\% of unsatisfied NPL of \textit{Op-O-3} because transparent architectures have a larger number of LPs due to smaller grooming capabilities. 
When the requested traffic is larger than 43Tb/s, the average unsatisfied NPL is 0 for all cases, indicating that the required NPL of all links in the network is satisfied. 
% the unsatisfied NPL of \textit{Tr-O-3} is up to 

\begin{figure}[htpb]
   \centering
        \includegraphics[width=0.75\linewidth]{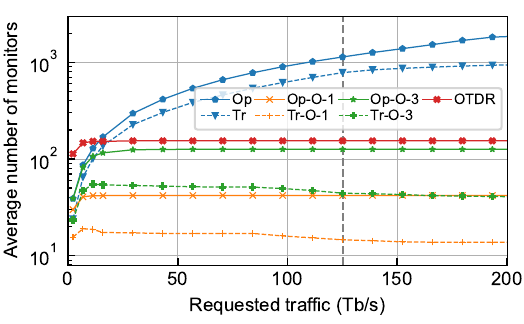}
    \caption{Average number of monitors vs. load in N14.}
    \label{fig:monitors-vs-load-nsf}
\end{figure}

Let us now evaluate the number of monitors under different loads in N14 as in Fig.~\ref{fig:monitors-vs-load-nsf}. The number of PPMs for \textit{Op} and \textit{Tr} always increases with the growing traffic due to the higher number of established LPs, and all LPs are equipped with PPMs. Note that OTDR modules are only placed to monitor the links that are monitored with PPM to have a fair comparison of OTDR and PPM. Therefore, as the traffic approaches 30Tb/s, the number of OTDRs increases to 154 and then remains constant, given that all links are already monitored by at least one PPM. 
% Thus, the number of OTDRs increases to xx when the traffic increases to xx Tb/s, and then remains constant since all the links are already monitored by at least one PPM. 
With \textit{OMP}, for \textit{Tr-O-1}, the number of monitors increases up to 19 when the traffic grows to 7Tb/s because more links are monitored. Then, the number of monitors decreases from 19 to 15 when the traffic grows because more LPs with a larger number of hops can be used. Instead, for \textit{Op-O-1}, the number of monitors never decreases since all LPs have only one hop. When NPL equals 3, the number of monitors of \textit{Op-O-3} and \textit{Tr-O-3} follows the consistent trend as in \textit{Op-O-1} and \textit{Tr-O-1}. 
% When \textit{NPL} equals to 3, the number of monitors of \textit{Op-O-3} increases up to xx when the requested traffic is xx Tb/s, indicating that the optimal solution of \textit{Op-O-3} has been achieved and all the links are monitored with \textit{NPL} equal to 3. In addition, the number of monitors in \textit{Tr-O-3} increases up to xx and then reduces to xx when the requested traffic increases. 
After all the links satisfy the required \textit{NPL}, the number of required OTDRs is greater than all the tested cases that place PPM with \textit{OMP}. For instance, the number of OTDRs required is around 7x to 11x the number of PPMs in \textit{Tr-O-1} when increasing the network traffic. 
%After describing the trends in the number of monitors, now let us compare the number of monitors under 1\% rejection probability. 

\begin{figure}[ht]%pb]
   \centering
        \includegraphics[width=0.75\linewidth]{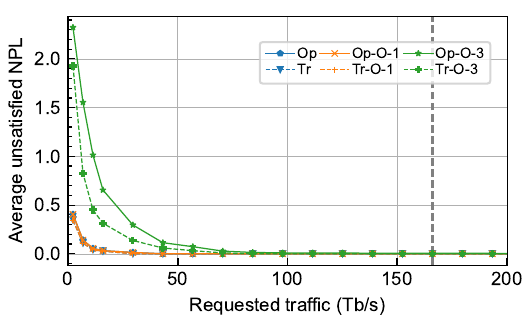}
    \caption{Average unsatisfied NPL vs. load in J14.}
    \label{fig:unsatisfied-NPL-Japan}
\end{figure}
\begin{figure}[htpb]
   \centering
        \includegraphics[width=0.75\linewidth]{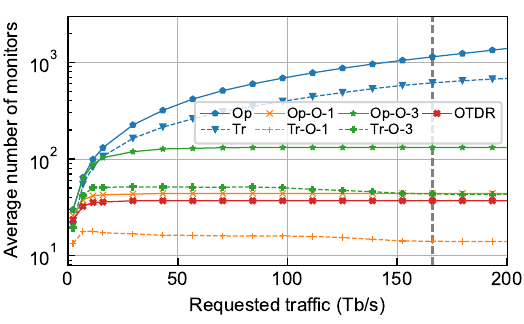}
    \caption{Average number of monitors vs. load in J14.}
    \label{fig:monitors-vs-load-japan}
\end{figure}

% \begin{figure*}[h]
%    \centering
%         \includegraphics[width=\linewidth]{cost_power_sensitivity_nsf.pdf}
%     \caption{Cost of and power consumption of different approaches in N14.}
%     \label{fig:cost-power-nsf}
% \end{figure*}

% \begin{figure*}[h]
%    \centering
%         \includegraphics[width=\linewidth]{cost_power_sensitivity_japan.pdf}
%     \caption{Cost of and power consumption of different approaches in J14.}
%     \label{fig:cost-power-japan}
% \end{figure*}

The unsatisfied NPL in J14 is shown in Fig.~\ref{fig:unsatisfied-NPL-Japan}. All the different approaches perform similarly to the unsatisfied NPL as in N14. Different from the results in N14, now there are no more unsatisfied NPL for all approaches when traffic reaches 99Tb/s (instead of 43Tb/s), because in J14, LPs can operate on higher modulation formats due to shorter link length, and hence fewer LPs are established to serve the same amount of traffic. Regarding the number of monitors, as shown in Fig.~\ref{fig:monitors-vs-load-japan}, all the approaches with PPM remain consistent trends as that in N14. 
% Specifically, all the links can be monitored with \textit{NPL} equal to 1 when the requested traffic exceeds xx Tb/s. Similarly, all the links can be monitored with \textit{NPL} equal to 3 when the requested traffic exceeds xy Tb/s. 
The difference is that, in J14, the number of OTDRs required to monitor the whole network reduces from 154 in N14 to 37 in J14, since J14 has a much shorter link length. Consequently, the number of OTDR required is smaller than all the cases with PPM except \textit{Tr-O-1}, as shown in Fig.~\ref{fig:monitors-vs-load-japan}.%, when the requested traffic is greater than 20Tb/s.
% For instance, the number of OTDRs required reduces from up to xx times the number of PPMs in N14 to xx times in J14 for \textit{Tr-O-1}. 

% \vspace{-3mm}
\begin{table}[h]
\small
\setlength\tabcolsep{3pt} 
\centering
\caption{Number of monitoring components}% on J14 and N14} 
\label{tab:results-monitors-1-percent}
\begin{tabular}{|c|c|c|c|c|c|c|c|}
\hline
\rowcolor[HTML]{C0C0C0} 
Scenarios                   & Op  & Tr  & Op-O-1 & Tr-O-1 & Op-O-3 & Tr-O-3 & OTDR \\ \hline
\cellcolor[HTML]{C0C0C0}J14 & 1167 & 624 & 44     & 14     & 132    & 43     & 37  \\ \hline
\cellcolor[HTML]{C0C0C0}N14 & 1135 & 782 & 42     & 15     & 126    & 44     & 154   \\ \hline
\end{tabular}
\end{table}

The analysis of the trends in the number of monitors indicates that, with optimized monitoring placement, the number of PPMs required has only a small variation with requested traffic unless under low-load traffic (e.g., less than 50 Tb/s, which corresponds to less than 30\% of spectrum occupation for transparent architecture). 
% the traffic is low (e.g., less than 50 Tb/s). 
Hence, in the following analysis, we compare the number of monitors under fixed traffic (i.e., traffic with 1\% rejection probability for transparent architecture, which corresponds to around 60\% and 70\% of spectrum occupation of J14 and N14, respectively). Note that, for the traffic with 1\% rejection probability, all the links can satisfy the required NPL for the tested cases. 
% As shown in Fig.~\ref{fig:monitors-vs-load-nsf} 
% The solid dotted line in Fig.~\ref{fig:monitors-vs-load-nsf} and Fig.~\ref{fig:monitors-vs-load-japan} represents the requested traffic when the network has 1\% rejection rate. 
The number of monitoring components (i.e., either PPM or OTDR modules) used in J14 and N14 topologies are reported in Table~\ref{tab:results-monitors-1-percent}. First, without \textit{OMP}, the number of PPM modules for \textit{Op} is about 30x the number of OTDR in J14, while this value is reduced to 7.4x in N14, as OTDR now experiences an about 300\% increase in N14 compared to J14 due to the fact that N14 has a much larger geographical dimension and hence much larger number of spans. With \textit{OMP}, the number of PPM modules for \textit{Op-O} is reduced to the number of edges in the network, as each edge is monitored with only one PPM module. 
% Note that in N14, the number of PPM modules is even decreased to 16\% of the number of OTDRs.
Note that, in N14, for \textit{Op-O-1}, the number of PPM modules even decreased to only 27\% of OTDRs. Moreover, \textit{Tr-O-1} can further reduce up to 65\% PPM modules by bypassing nodes. Specifically, the number of OTDR is about 10x the number of PPM, suggesting that the cost (power) of one PPM should be 10x lower than that of OTDR to be a competing technology. For J14, the number of OTDR is 2.6x the number of PPM due to the shorter link length, suggesting that the cost (power) of one PPM should be 2.6x lower than that of OTDR. When the required NPL is 3, the number of required monitors is around 3 times that of cases when the required NPL is equal to 1. 
% Instead, the number of OTDRs only depends on network topology (the links to monitor) rather than network architectures. 

% \begin{figure*}[h]
%    \centering
%         \includegraphics[width=0.92\linewidth]{cost_crossing_point_sensitivity_nsf.pdf}
%     \caption{Crossing value of cost of and power consumption under different NPL in N14.}
%     \label{fig:crossing-value-nsf}
% \end{figure*}

% \begin{figure*}[h]
%    \centering
%         \includegraphics[width=0.92\linewidth]{cost_crossing_point_sensitivity_japan.pdf}
%     \caption{Crossing value of cost of and power consumption of different NPL in J14.}
%     \label{fig:crossing-value-japan}
% \end{figure*}

\subsection{Evaluation of Cost and Power Consumption}
We now evaluate the cost and power consumption of PPM compared to OTDR in N14 and J14.% as shown in Fig.~\ref{fig:cost-power-nsf} and Fig.~\ref{fig:cost-power-japan}.

% \subsubsection{Cost and Power Consumption in N14}

% First, without \textit{OMP}, \textit{Tr} architecture reduces the number of PPM modules by 36\% compared to \textit{Op} for J14, which is decreased to 19\% for N14 as less nodes can be bypassed in a larger topology. With \textit{OMP}, the number of PPM modules required for \textit{Op-O} are exactly the number of edges in the network as one edge is monitored with one PPM module. When bypassing is allowed, \textit{Tr-O} reduces around 60\% of PPM modules for both topologies because LPs for requests with small data rates can cover a similar number of edges due to the long reach of signals. 
% % Without \textit{OMP}, compared to J14, N14 requires 8\% additional PPM for \textit{Op} and 36\% additional PPM for \textit{Tr} due to the larger topology and more intermediate nodes with regeneration. However, with OMP, the number of PPMs in both topologies is almost the same, as requests with small data rates can cover a similar number of edges due to the long reach of signals. 
% Notably, the OTDR in N14 shows an 832\% increase compared to J14, owing to the increased number of spans in each edge.

\begin{figure*}[!h]
   \centering
        \includegraphics[width=0.7\linewidth]{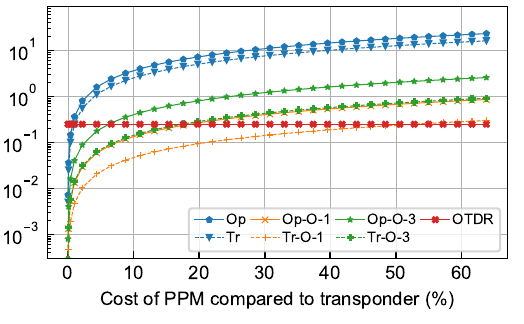}
    \caption{Cost of different approaches in N14.}
    \label{fig:cost-power-nsf-cost}
\end{figure*}
\begin{figure*}[!h]
   \centering
        \includegraphics[width=0.7\linewidth]{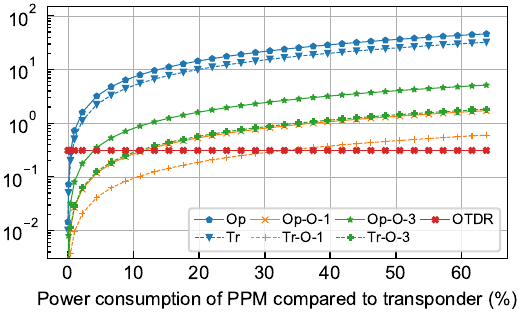}
    \caption{Power consumption of different approaches in N14.}
    \label{fig:cost-power-nsf-power}
\end{figure*}

{\em Normalized cost in N14.} Fig.~\ref{fig:cost-power-nsf-cost} illustrates the normalized cost of monitoring per Tb/s for N14. We define the x-axis value of the point where PPM's cost/power consumption is equal to OTDR as \textit{crossing value}, which is also reported in Table~\ref{tab:crossing-value-nsf}. The cases with \textit{OMP} have much smaller crossing values than the cases without \textit{OMP}, showing that \textit{OMP} is fundamental to lower PPM monitoring costs. Without \textit{OMP}, the cost of one PPM must be less than 0.7\% and 1.0\% of one transponder for \textit{Op} and \textit{Tr}, respectively, to compete with OTDR. However, with \textit{OMP}, the cost of PPM should be below 18.3\% and 52.7\% of the cost of transponder for \textit{Op-O-1} and \textit{Tr-O-1}, respectively, to compete with OTDR. 
Note that the cost difference between \textit{Op-O-1} and \textit{Tr-O-1} is due to the bypassing of nodes in the transparent architecture. 
Moreover, when the required NPL is 3, the crossing value of cost is reduced to around 1/3 of that of the cases with the required NPL equal to 1. Specifically, the crossing value of \textit{Op-O-3} and \textit{Tr-O-3} is 6.1\% and 17.4\%, respectively. 

\begin{table}[htp]
\small
\setlength\tabcolsep{3.5pt} 
\centering
\caption{Crossing values of PPM with OTDR in N14} 
\begin{tabular}{|c|c|c|c|c|c|c|}
\hline
\rowcolor[HTML]{C0C0C0} 
Scenarios &
  Op &
  Tr &
  Op-O-1 &
  Tr-O-1 &
  \multicolumn{1}{l|}{\cellcolor[HTML]{C0C0C0}Op-O-3} &
  \multicolumn{1}{l|}{\cellcolor[HTML]{C0C0C0}Tr-O-3} \\ \hline
\rowcolor[HTML]{FFFFFF} 
\cellcolor[HTML]{C0C0C0}Cost &
  0.7 &
  1.0 &
  18.3 &
  52.7 &
  6.1 &
  17.4 \\ \hline
\rowcolor[HTML]{FFFFFF} 
\cellcolor[HTML]{C0C0C0}Power consumption &
  0.4 &
  0.6 &
  11.5 &
  33.0 &
  3.8 &
  10.9 \\ \hline
\end{tabular}
\label{tab:crossing-value-nsf}
\end{table}
% demonstrates that if we do not optimize the monitoring placement for PPM, its cost of monitoring is lower than OTDR only when the cost of one PPM compared to one transponder is lower than approximately 0.2\% and 0.3\% for \textit{Op} and \textit{Tr}, respectively, which requires that PPM has very low complexity. 
%Note that the elevated cost of PPM without \textit{OMP} compared to OTDR is due to an increase in the number of PPM with traffic.
% The reason why we have such a high cost of monitoring for OTDR is that the number of monitoring components of PPM increases with traffic.
% The reason why the cost of transparent architecture is lower than opaque architecture is that some nodes are bypassed. 

{\em Normalized power consumption in N14.}  Fig.~\ref{fig:cost-power-nsf-power} plots the power consumption of PPM compared to OTDR. 
% To consume less power than OTDR, PPM needs to have lower complexity. 
Without \textit{OMP}, the power consumption of PPM relative to one transponder must be less than 0.4\% and 0.6\% of the transponder's power consumption. 
% that of the transponder to have a lower power consumption than OTDR.
With \textit{OMP}, the power consumption of PPM can be up to 11.5\% and 33.0\% of the transponder's power consumption for \textit{Op-O-1} and \textit{Tr-O-1}, respectively. When the required NPL is 3, the crossing value of power consumption is reduced to around 1/3 that of the cases when the required NPL is 1.

% \begin{table}[h]
% \small
% \setlength\tabcolsep{8pt} 
% \centering
% \caption{Crossing values of PPM with OTDR in N14.} 
% \begin{tabular}{|c|c|c|c|c|}
% \hline
% \rowcolor[HTML]{C0C0C0} 
% Scenarios                          & Op       & Tr       & Op-O   & Tr-O   \\ \hline
% \cellcolor[HTML]{C0C0C0}Cost & 1.5(0.2) & 2.0(0.4) & 33(4) & 80(12) \\ \hline
% \cellcolor[HTML]{C0C0C0}Power consumption & 0.9(0.1) & 1.3(0.2) & 20(3) & 50(8) \\ \hline
% \end{tabular}
% \label{tab:crossing-point}
% \end{table}

% \subsubsection{Cost and Power Consumption in J14}

{\em Normalized cost and power consumption in J14.}  The overall trends of cost and power consumption in J14 are similar to N14. Table~\ref{tab:crossing_value_japan} reports the crossing values for J14. Different from N14, PPM can perform better than OTDR only with a much lower cost and power consumption than one transponder. This is because less OTDRs are deployed in links in a smaller topology. 
% , resulting in lower cost and power consumption of OTDRs. 
Specifically, PPM's cost and power consumption are reduced to approximately 0.2\% and 0.1\% for \textit{Op}, and 0.3\% and 0.2\% for \textit{Tr}, respectively. Moreover, with \textit{OMP}, the crossing values of cost and power consumption are also significantly reduced. For instance, the crossing values of cost and power consumption for \textit{Tr-O-1} are reduced from 52.7\% to 13.2\% and 33.0\% to 8.3\% for transparent architecture.

\begin{table}[h]
\small
\setlength\tabcolsep{3.5pt} 
\centering
\caption{Crossing values of PPM with OTDR in J14} 
\begin{tabular}{|c|c|c|c|c|c|c|}
\hline
\rowcolor[HTML]{C0C0C0} 
Scenarios &
  Op &
  Tr &
  Op-O-1 &
  Tr-O-1 &
  \multicolumn{1}{l|}{\cellcolor[HTML]{C0C0C0}Op-O-3} &
  \multicolumn{1}{l|}{\cellcolor[HTML]{C0C0C0}Tr-O-3} \\ \hline
\rowcolor[HTML]{FFFFFF} 
\cellcolor[HTML]{C0C0C0}Cost &
  0.2 &
  0.3 &
  4.2 &
  13.2 &
  1.4 &
  4.3 \\ \hline
\rowcolor[HTML]{FFFFFF} 
\cellcolor[HTML]{C0C0C0}Power consumption &
  0.1 &
  0.2 &
  2.6 &
  8.3 &
  0.9 &
  2.7 \\ \hline
\end{tabular}
\label{tab:crossing_value_japan}
\end{table}

\section{Conclusion}
\label{sec:conclusion}

Following the recent experimentation of PPM as a novel monitoring solution, in this work, we investigated the problem of optimized monitoring placement for PPM, and we quantified the cost and power consumption of PPM compared to OTDR to provide guidelines for deployment of PPM modules. 
Results show that, even with the promising reduction of cost and power consumption using PPM for optical monitoring, for a nation-wide topology as J14, cost and power consumption of the PPM should be below 2.6 times that of OTDR. This requirement enforces that the PPM's cost needs to be less than 13\% of that of a transponder, and PPM's power needs to be less than 8\% of that of a transponder. 
Instead, in a larger continental-wide topology as N14, both PPM's cost and power consumption should be below 10.2 times that of OTDR, which enforces that the PPM's cost needs to be less than 53\% of that of a transponder, and PPM's power needs to be less than 33\% of that of a transponder. 
% Results show that,
% % Even with the promising reduction of cost and power consumption using PPM for optical monitoring, 
% for a nation-wide topology as J14, 
% cost (power consumption) of the PPM should be below 2.6 times that of OTDR, which corresponds to 13\% (8\%) of that of a transponder.
% %to compete with OTDR. 
% % in terms of cost (energy consumption). 
% Instead, in a larger continental-wide topology as N14, 
% % cost and power consumption of PPM are not limiting factors for PPM deployments, as PPM is not expected to have a high cost and power consumption above 76\% (47\%) of that of a transponder.
% PPM's cost (power consumption) should be below 10.2 times that of OTDR, which corresponds to 53\% (33\%) of that of a transponder. 
This means that the cost and power consumption of PPM are not limiting factors for PPM deployments, as PPM is not expected to have such a high cost and power consumption. In the future, we plan to deploy PPMs in the field to evaluate their cost and power consumption further.

\section*{Acknowledgment}
This work was supported by the Italian Ministry of University and Research (MUR) and the European Union (EU) under the PON/REACT project.
It is also funded by the EU Horizon
2020 B5G-OPEN Project (101016). 
Moreover, this work was supported by the European Union -Next Generation EU under the Italian National Recovery and Resilience Plan (NRRP), Mission 4, Component 2, Investment 1.3, CUP D43C22003080001, partnership on “Telecommunications of the Future” (PE00000001 - program “RESTART”).

% and the European Union under the Italian National Recovery and Resilience Plan (NRRP) of NextGenerationEU, partnership on “Telecommunications of the Future” (PE00000001 - program “RESTART”). 

% \appendix
% \section{Example Appendix Section}

 \bibliographystyle{elsarticle-num} 
 \bibliography{reference}

\begin{thebibliography}{10}
\expandafter\ifx\csname url\endcsname\relax
  \def\url#1{\texttt{#1}}\fi
\expandafter\ifx\csname urlprefix\endcsname\relax\def\urlprefix{URL }\fi
\expandafter\ifx\csname href\endcsname\relax
  \def\href#1#2{#2} \def\path#1{#1}\fi

\bibitem{he2023improved}
Y.~He, Z.~Zhai, L.~Dou, L.~Wang, Y.~Yan, C.~Xie, C.~Lu, A.~P.~T. Lau, Improved qot estimations through refined signal power measurements and data-driven parameter optimizations in a disaggregated and partially loaded live production network, IEEE/Optica Journal of Optical Communications and Networking 15~(9) (2023) 638--648.

\bibitem{wang2024digital}
R.~Wang, J.~Zhang, Z.~Gu, M.~Ibrahimi, B.~Zhang, F.~Musumeci, M.~Tornatore, Y.~Ji, Digital-twin-assisted meta learning for soft-failure localization in roadm-based optical networks, IEEE/Optica Journal of Optical Communications and Networking 16~(7) (2024) C11--C19.

\bibitem{vejdannik2023leveraging}
M.~Vejdannik, A.~Sadr, Leveraging genetic algorithm to address multi-failure localization in optical networks, Optical Switching and Networking 47 (2023) 100706.

\bibitem{pang2024large}
Y.~Pang, M.~Zhang, Y.~Liu, X.~Li, Y.~Wang, Y.~Huan, Z.~Liu, J.~Li, D.~Wang, Large language model-based optical network log analysis using llama2 with instruction tuning, IEEE/Optica Journal of Optical Communications and Networking 16~(11) (2024) 1116--1132.

\bibitem{wang2025multi}
R.~Wang, Q.~Zhang, J.~Zhang, Z.~Gu, M.~Ibrahimi, H.~Yu, B.~Zhang, F.~Musumeci, Y.~Ji, M.~Tornatore, Multi-failure localization in high-degree {ROADM}-based optical networks using rules-informed neural networks, IEEE Journal on Selected Areas in Communications 43~(5) (2025) 1738--1754.

\bibitem{zhou2016field}
Y.~R. Zhou, K.~Smith, P.~Weir, A.~Lord, J.~Chen, W.~Pan, N.~Zhou, Z.~Xiao, Field trial demonstration of novel optical superchannel capacity protection for 400g using dp--16qam and dp--qpsk with in-service otdr fault localization, in: Optical Fiber Communication Conference (OFC), 2016.

\bibitem{anderson2004troubleshooting}
D.~R. Anderson, L.~M. Johnson, F.~G. Bell, Troubleshooting optical fiber networks: understanding and using optical time-domain reflectometers, Elsevier, 2004.

\bibitem{BrianSuboptic}
B.~Jander, L.~Garrett, R.~Kram, L.~Richardson, Y.~Xu, M.~Rodriguez, D.~Pappas, Y.~Jiang, Y.~Chai, T.~Verdi, J.~Giotis, F.~Ayadi, Enhanced undersea line monitoring technology for coherent open cable systems, in: SubOptic Conference, 2019.

\bibitem{Hahn:22}
C.~Hahn, J.~Chang, Z.~Jiang, Monitoring of generalized optical signal-to-noise ratio using in-band spectral correlation method, in: European Conference on Optical Communication (ECOC), 2022.

\bibitem{Sasai:23}
T.~Sasai, Y.~Sone, E.~Yamazaki, M.~Nakamura, Y.~Kisaka, A generalized method for fiber-longitudinal power profile estimation, in: European Conference on Optical Communication (ECOC), 2023.

\bibitem{maySubmine}
A.~May, F.~Boitier, A.~C. Meseguer, J.~U. Esparza, P.~Plantady, A.~Calsat, P.~Layec, Longitudinal power monitoring over a deployed 10,000-km link for submarine systems, in: Optical Fiber Communication Conference (OFC), 2023.

\bibitem{tanimura2020fiber}
T.~Tanimura, S.~Yoshida, K.~Tajima, S.~Oda, T.~Hoshida, Fiber-longitudinal anomaly position identification over multi-span transmission link out of receiver-end signals, Journal of Lightwave Technology 38~(9) (2020) 2726--2733.

\bibitem{may2021receiver}
A.~May, F.~Boitier, E.~Awwad, P.~Ramantanis, M.~Lonardi, P.~Ciblat, Receiver-based experimental estimation of power losses in optical networks, IEEE Photonics Technology Letters 33~(22) (2021) 1238--1241.

\bibitem{sasaiLinear}
T.~Sasai, M.~Takahashi, M.~Nakamura, E.~Yamazaki, Y.~Kisaka, Linear least squares estimation of fiber-longitudinal optical power profile, Journal of Lightwave Technology 42~(6) (2024) 1955--1965.
\newblock \href {https://doi.org/10.1109/JLT.2023.3327760} {\path{doi:10.1109/JLT.2023.3327760}}.

\bibitem{KimPPM}
I.~Kim, O.~Vassilieva, R.~Shinzaki, M.~Eto, S.~Oda, P.~Palacharla, Multi-channel longitudinal power profile estimation, in: European Conference on Optical Communication (ECOC), 2023.

\bibitem{OTDRref}
{Cisco}, {Cisco Network Convergence System 1001 OTDR Line Card Data Sheet}, \url{https://www.cisco.com/c/en/us/products/collateral/optical-networking/network-convergence-system-1000-series/datasheet-c78-742294.html/}, accessed on 18/02/2024.

\bibitem{OTDRRefRange}
{Kingfisherfiber}, {Specifications for OTDR Handheld Optical Time Domain Reflectometer}, \url{https://kingfisherfiber.com/media/2246/6700-otdr.pdf}, accessed on 18/02/2024.

\bibitem{mayDemonstration}
A.~May, F.~Boitier, A.~Courilleau, B.~Al~Ayoubi, P.~Layec, Demonstration of enhanced power losses characterization in optical networks, in: Optical Fiber Communication Conference (OFC), 2022, pp. Th1C--6.

\bibitem{zhang2024power}
Q.~Zhang, A.~Morea, P.~Layec, M.~Ibrahimi, F.~Musumeci, M.~Tornatore, Power-consumption analysis for different ipowdm network architectures with zr/zr+ and long-haul muxponders, Journal of Optical Communications and Networking 16~(12) (2024) 1189--1203.

\bibitem{Qiaolun:23}
Q.~Zhang, P.~Layec, A.~Morea, M.~Tornatore, Cost and power-consumption analysis for power profile monitoring in optical networks, in: European Conference on Optical Communication (ECOC), 2023.

\bibitem{sasaiPerformance}
T.~Sasai, E.~Yamazaki, Y.~Kisaka, Performance limit of fiber-longitudinal power profile estimation methods, Journal of Lightwave Technology 41~(11) (2023) 3278--3289.

\bibitem{yang2024integrating}
X.~Yang, T.~Eldahrawy, C.~Sun, A.~Lorences-Riesgo, M.~Tornatore, G.~Charlet, Y.~Pointurier, Integrating {PPE} and inputs refinement for enhanced qot estimation and optimization in an optical mesh network, in: European Conference on Optical Communication (ECOC), VDE, 2024, pp. 491--494.

\bibitem{may2025accuracy}
A.~May, F.~Boitier, P.~Layec, Accuracy assessment of power profile estimation using {MMSE} or deconvoluted profiles, Journal of Lightwave Technology, early access (2025).
\newblock \href {https://doi.org/10.1109/JLT.2025.3565167} {\path{doi:10.1109/JLT.2025.3565167}}.

\bibitem{sasaiDigital}
T.~Sasai, M.~Nakamura, E.~Yamazaki, S.~Yamamoto, H.~Nishizawa, Y.~Kisaka, Digital longitudinal monitoring of optical fiber communication link, Journal of Lightwave Technology 40~(8) (2022) 2390--2408.

\bibitem{sena2022advanced}
M.~Sena, P.~Hazarika, C.~Santos, B.~Correia, R.~Emmerich, B.~Shariati, A.~Napoli, V.~Curri, W.~Forysiak, C.~Schubert, et~al., Advanced {DSP}-based monitoring for spatially resolved and wavelength-dependent amplifier gain estimation and fault location in {C+L}-band systems, Journal of Lightwave Technology 41~(3) (2022) 989--998.

\bibitem{may2022receiver}
A.~May, E.~Awwad, P.~Ramantanis, P.~Ciblat, Receiver-based localization and estimation of polarization dependent loss, in: OptoElectronics and Communications Conference (OECC), IEEE, 2022, pp. 1--4.

\bibitem{andrenacciPdl}
L.~Andrenacci, G.~Bosco, D.~Pilori, {PDL} localization and estimation through linear least squares-based longitudinal power monitoring, IEEE Photonics Technology Letters 35~(24) (2023) 1431--1434.

\bibitem{10484838}
M.~Eto, K.~Tajima, K.~Sone, S.~Yoshida, R.~Shinzaki, S.~Oda, T.~Hoshida, Fibre type identification based on power profile estimation, in: European Conference on Optical Communications (ECOC), Vol. 2023, 2023, pp. 127--130.

\bibitem{hahnLocalization}
C.~Hahn, J.~Chang, Z.~Jiang, Localization of reflection induced multi-path-interference over multi-span transmission link by receiver-side digital signal processing, in: Optical Fiber Communication Conference, 2022.

\bibitem{10527091}
C.~Hahn, J.~Chang, Z.~Jiang, Estimation and localization of {DGD} distributed over multi-span optical link by correlation template method, in: Optical Fiber Communications Conference and Exhibition (OFC), 2024, pp. 1--3.

\bibitem{10810103}
C.~Hahn, J.~Chang, Z.~Jiang, Location-resolved {gOSNR} estimation of multi-span optical transmission systems, in: Asia Communications and Photonics Conference (ACP), 2024, pp. 1--3.

\bibitem{10526939}
I.~Kim, K.~Sone, O.~Vassilieva, S.~Oda, P.~Palacharla, T.~Hoshida, Nonlinear {SNR} estimation based on power profile estimation in hybrid {Raman}-{EDFA} link, in: Optical Fiber Communications Conference and Exhibition (OFC), 2024, pp. 1--3.

\bibitem{delezoideField}
C.~Delezoide, P.~Ramantanis, L.~Gifre, F.~Boitier, P.~Layec, Field trial of failure localization in a backbone optical network, in: European Conference on Optical Communication (ECOC), 2021.

\bibitem{angelou2012optimized}
M.~Angelou, Y.~Pointurier, D.~Careglio, S.~Spadaro, I.~Tomkos, Optimized monitor placement for accurate qot assessment in core optical networks, IEEE/Optica Journal of Optical Communications and Networking 4~(1) (2012) 15--24.

\bibitem{christodoulopoulos2016exploiting}
K.~Christodoulopoulos, N.~Sambo, E.~Varvarigos, Exploiting network kriging for fault localization, in: Optical fiber communication conference (OFC), 2016.

\bibitem{ZhangProgressive22}
Q.~Zhang, O.~Ayoub, J.~Wu, F.~Musumeci, G.~Li, M.~Tornatore, Progressive slice recovery with guaranteed slice connectivity after massive failures, IEEE/ACM Transactions on Networking 30~(2) (2022) 826--839.

\bibitem{arrigoni2023tomography}
V.~Arrigoni, M.~Prata, N.~Bartolini, Tomography-based progressive network recovery and critical service restoration after massive failures, in: IEEE INFOCOM 2023-IEEE Conference on Computer Communications, IEEE, 2023, pp. 1--10.

\bibitem{machuca2004optimal}
C.~M. Machuca, I.~Tomkos, Optimal monitoring equipment placement for fault and attack location in transparent optical networks, in: International Conference on Research in Networking, Springer, 2004, pp. 1395--1400.

\bibitem{thiruvasagam2021reliable}
P.~K. Thiruvasagam, A.~Chakraborty, A.~Mathew, C.~S.~R. Murthy, Reliable placement of service function chains and virtual monitoring functions with minimal cost in softwarized 5g networks, IEEE Transactions on Network and Service Management 18~(2) (2021) 1491--1507.

\bibitem{li2024drl}
M.~Li, Q.~Zhang, A.~Gatto, S.~Bregni, G.~Verticale, M.~Tornatore, Drl-based progressive recovery for quantum-key-distribution networks, IEEE/Optica Journal of Optical Communications and Networking 16~(9) (2024) E36--E47.

\bibitem{wang2024alarmgpt}
Y.~Wang, C.~Zhang, J.~Li, Y.~Pang, L.~Zhang, M.~Zhang, D.~Wang, Alarmgpt: an intelligent alarm analyzer for optical networks using a generative pre-trained transformer, IEEE/Optica Journal of Optical Communications and Networking 16~(6) (2024) 681--694.

\bibitem{machuca2007optimal}
C.~M. Machuca, M.~Kiese, Optimal placement of monitoring equipment in transparent optical networks, in: 2007 6th International Workshop on Design and Reliable Communication Networks, IEEE, 2007, pp. 1--6.

\bibitem{abrams2004set}
Z.~Abrams, A.~Goel, S.~Plotkin, Set k-cover algorithms for energy efficient monitoring in wireless sensor networks, in: Proceedings of the 3rd international symposium on Information processing in sensor networks, 2004, pp. 424--432.

\bibitem{zamiReachTable}
T.~Zami, B.~Lavigne, M.~Lefran{\c{c}}ois, Added value of 90 gbaud transponders for wdm networks, in: Optical Fiber Communications Conference and Exhibition (OFC), 2020.

\bibitem{slijepcevic2001power}
S.~Slijepcevic, M.~Potkonjak, Power efficient organization of wireless sensor networks, in: IEEE International Conference on Communications (ICC), 2001.

\bibitem{vasko2016best}
F.~J. Vasko, Y.~Lu, K.~Zyma, What is the best greedy-like heuristic for the weighted set covering problem?, Operations Research Letters 44~(3) (2016) 366--369.

\bibitem{bestuzheva2023enabling}
K.~Bestuzheva, M.~Besan{\c{c}}on, W.-K. Chen, A.~Chmiela, T.~Donkiewicz, J.~van Doornmalen, L.~Eifler, O.~Gaul, G.~Gamrath, A.~Gleixner, et~al., Enabling research through the scip optimization suite 8.0, ACM Transactions on Mathematical Software 49~(2) (2023) 1--21.

\bibitem{gabriel1969new}
K.~R. Gabriel, R.~R. Sokal, A new statistical approach to geographic variation analysis, Systematic zoology 18~(3) (1969) 259--278.

\bibitem{velinska2017optical}
J.~Velinska, M.~Mirchev, I.~Mishkovski, Optical networks’ topologies: costs, routing and wavelength assignment, Optical Networks (2017) 10.

\bibitem{ibrahimi2021machine}
M.~Ibrahimi, H.~Abdollahi, C.~Rottondi, A.~Giusti, A.~Ferrari, V.~Curri, M.~Tornatore, Machine learning regression for qot estimation of unestablished lightpaths, IEEE/Optica Journal of Optical Communications and Networking 13~(4) (2021) B92--B101.

\bibitem{bathula2009constraint}
B.~G. Bathula, J.~M. Elmirghani, Constraint-based anycasting over optical burst switched networks, IEEE/Optica Journal of Optical Communications and Networking 1~(2) (2009) A35--A43.

\end{thebibliography}




% %% For numbered reference style

%   \textit{\LaTeX: a document preparation system},
%   Addison Wesley, Massachusetts,

% \end{thebibliography}
\end{document}